\documentclass[%
 reprint,
 amsmath,amssymb,
 aps,superscriptaddress,
]{revtex4-2}

\usepackage{graphicx}
\usepackage{dcolumn}
\usepackage{bm}
\usepackage{braket}
\usepackage{xfrac}
\usepackage[separate-uncertainty = true, multi-part-units=single]{siunitx}
\usepackage{xcolor}
\usepackage{gensymb}
\usepackage{natbib}
\usepackage[english]{babel}

\begin{document}

\preprint{APS/123-QED}

\title{On-chip quantum interference between independent lithium niobate-on-insulator photon-pair sources}

\author{Robert J. Chapman}
\email{rchapman@ethz.ch}
\affiliation{Optical Nanomaterial Group, Institute for Quantum Electronics, Department of Physics, ETH Zurich, CH-8093 Zurich, Switzerland}

\author{Tristan Kuttner}
\affiliation{Optical Nanomaterial Group, Institute for Quantum Electronics, Department of Physics, ETH Zurich, CH-8093 Zurich, Switzerland}

\author{Jost Kellner}
\affiliation{Optical Nanomaterial Group, Institute for Quantum Electronics, Department of Physics, ETH Zurich, CH-8093 Zurich, Switzerland}

\author{Alessandra Sabatti}
\affiliation{Optical Nanomaterial Group, Institute for Quantum Electronics, Department of Physics, ETH Zurich, CH-8093 Zurich, Switzerland}

\author{Andreas Maeder}
\affiliation{Optical Nanomaterial Group, Institute for Quantum Electronics, Department of Physics, ETH Zurich, CH-8093 Zurich, Switzerland}

\author{Giovanni Finco}
\affiliation{Optical Nanomaterial Group, Institute for Quantum Electronics, Department of Physics, ETH Zurich, CH-8093 Zurich, Switzerland}

\author{Fabian Kaufmann}
\affiliation{Optical Nanomaterial Group, Institute for Quantum Electronics, Department of Physics, ETH Zurich, CH-8093 Zurich, Switzerland}

\author{Rachel Grange}
\affiliation{Optical Nanomaterial Group, Institute for Quantum Electronics, Department of Physics, ETH Zurich, CH-8093 Zurich, Switzerland}

\date{\today}

\begin{abstract}
Generating and interfering non-classical states of light is fundamental to optical quantum information science and technology.
Quantum photonic integrated circuits provide one pathway towards scalability by combining nonlinear sources of non-classical light and programmable circuits in centimeter-scale devices.
The key requirements for quantum applications include efficient generation of indistinguishable photon-pairs and high-visibility programmable quantum interference.
Here, we demonstrate a lithium niobate-on-insulator (LNOI) integrated photonic circuit that generates a two-photon path-entangled state, and a programmable interferometer for quantum interference.
We generate entangled photons with $\sim\SI{2.3e8}{pairs/s/mW}$ brightness and perform quantum interference experiments on the chip with \SI{96.8(3.6)}{\percent} visibility.
LNOI is an emerging photonics technology that has revolutionized high-speed modulators and efficient frequency conversion.
Our results provide a path towards large-scale integrated quantum photonics including efficient photon-pair generation and programmable circuits for applications such as boson sampling and quantum communications.
\end{abstract}

\maketitle

Optical quantum computers utilize non-classical light sources, interferometric circuits and single photon detection to perform tasks such as boson sampling that can out-perform even world-leading supercomputers \cite{zhong_quantum_2020, madsen_quantum_2022, deng_gaussian_2023}.
Research into quantum photonic integrated circuits has witnessed significant progress in the last years \cite{wang_integrated_2020}.
These devices miniaturize table-top experiments onto centimeter-scale chips that can run more efficiently, have the advantage of stability, and have the potential to scale far beyond their free-space optics counterparts.
Today's leading integrated quantum photonics experiments use silicon and silicon-nitride waveguides that benefit from well-established nanofabrication processes and widely available foundry services for small to medium scale experiments \cite{adcock_advances_2021}.
While silicon and silicon-nitride photonics can generate photon-pairs by nonlinear four-wave-mixing \cite{arrazola_quantum_2021, bao_very-large-scale_2023}, this process is inefficient compared to three-wave spontaneous parametric down-conversion (SPDC) in $\chi^{(2)}$ crystals such as lithium niobate and potassium titanyl phosphate.
However, these crystals are more challenging to fabricate into scalable quantum photonic integrated circuits.
As a result, the leading quantum experiments to-date still use free-space SPDC sources and circuits to prepare entangled photons \cite{zhong_12-photon_2018, konno_logical_2024} and perform post-classical boson sampling \cite{zhong_quantum_2020, madsen_quantum_2022, deng_gaussian_2023}.

Lithium niobate integrated photonics is seeing a resurgence with the commercial availability of thin-film lithium-niobate-on-insulator (LNOI) wafers that can be used for high index contrast waveguides with extremely high nonlinear and electro-optic coefficients compared with bulk material \cite{zhu_integrated_2021}.
LNOI photonics has made tremendous progress towards high-speed optical modulators \cite{wang_integrated_2018} and ultra-efficient frequency conversion with periodically poled waveguides \cite{wang_ultrahigh-efficiency_2018, lu_periodically_2019}
Despite its potential for quantum photonics, there is little progress in applying LNOI for quantum information processing beyond demonstrations of photon-pair generation \cite{zhao_high_2020, xin_spectrally_2022} and interference of externally generated photons \cite{chapman_quantum_2023, babel_demonstration_2023}.
One of the key requirements for optical quantum information processing is the generation of indistinguishable photons and high-visibility interference  \cite{silverstone_-chip_2014}.
Photon indistinguishability is crucial for generating entanglement with linear optical quantum computing \cite{knill_scheme_2001, ralph_linear_2002}, and high-visibility interference is necessary for performing high-fidelity quantum gates \cite{taballione_20-mode_2023}.
Integrating photon-pair generation and programmable interferometers on a single LNOI device represents a major step in quantum photonics towards and achieving an all-in-chip quantum advantage.

Here, we demonstrate a LNOI quantum photonic chip that generates indistinguishable photons in a path-entangled state, and has a programmable circuit for performing tunable quantum interference.
We generate a two-photon $N00N$ state via SPDC in two identical periodically poled LNOI waveguides, and interfere the photons with an on-chip interferometer.
By modifying the phase of the $N00N$ state and the interferometer phase, we observe up to \SI{96.8(3.6)}{\percent} visibility quantum interference that produces bunched or anti-bunched photons.
We generate photon-pairs with an on-chip brightness of $\sim\SI{2.3e8}{pairs/s/mW}$, which is significantly higher than bulk-optics source, and verify the output two-photon state in an external Hong-Ou-Mandel experiment. 
Our work combines state-of-the-art quantum light sources and programmable circuits in a single device that paves the way towards large-scale quantum photonics with LNOI technology.

\begin{figure*}
    \centering
    \includegraphics[width=1.0\linewidth]{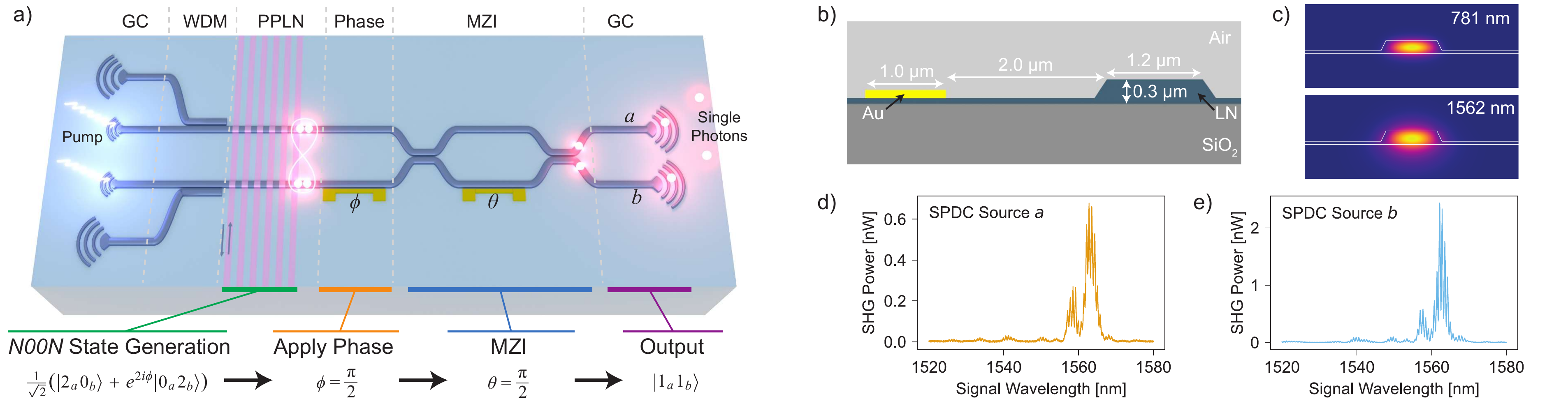}
    \caption{\textbf{LNOI quantum photonic chip.} 
    a) The pump laser generates a path-entangled $N00N$ state across two periodically poled LNOI waveguides.
    The $N00N$ state phase $\phi$ is controlled with a thermo-optic phase shifter, and a MZI acts as a tunable beamsplitter to enable quantum interference. GC: Grating coupler, WDM: Wavelength division multiplexer, PPLN: Periodically poled LNOI, MZI: Mach-Zehnder interferometer. b) Cross section of the LNOI waveguide and gold thermo-optic phase shifter. 
    c) Optical mode at \SI{781}{nm} and \SI{1562}{nm}. 
    d) \& e) Second harmonic generation in the periodically poled LNOI waveguide $a$ and $b$.
    }
    \label{fig:schematic}
\end{figure*}

\section*{Results}
Quantum photonic processors rely on Hong-Ou-Mandel (HOM) interference of indistinguishable photons to generate entangled states for linear optical quantum computing \cite{hong_measurement_1987, knill_scheme_2001, ralph_linear_2002}.
The HOM effect describes the bunching of photons at a 50:50 beamsplitter, which generates a ``$N00N$ state'' -- a superposition of $N$ photons in the first mode and $N$ photons in the second mode -- with $N=2$ for the case of two input photons such as those generated by SPDC.
The $N00N$ state can be written as
\begin{equation}
    \ket{\psi_{N00N}} = \tfrac{1}{\sqrt{2}}(\ket{N_a0_b} + e^{iN\phi}\ket{0_aN_b}),
    \label{eqn:N00N_}
\end{equation}
where $a$ and $b$ indicate the mode, and $\phi$ is the phase with an additional factor $N$ as a consequence of having a superposition of $N$ photons in each waveguide. 
If we reverse the experiment and inject a $N00N$ state into a beamsplitter the photons will either bunch or anti-bunch depending on the phase that is $N$ times more sensitive than classical interference, which is one of the key principles of quantum metrology \cite{dowling_quantum_2008}.
This phenomenon can also be used to separate indistinguishable photons generated by nonlinear sources into two distinct modes \cite{silverstone_-chip_2014, jin_-chip_2014, bao_very-large-scale_2023}.

Our LNOI quantum interference device is illustrated in Figure \ref{fig:schematic}a.
We coherently pump two periodically poled LNOI waveguides with a continuous-wave laser at \SI{781}{nm} wavelength that generate photon-pairs centered at \SI{1562}{nm} via type-0 SPDC.
By pumping the two SPDC sources, we generate the path-entangled two-photon $N00N$ state
\begin{equation}
    \ket{\psi} = \tfrac{1}{\sqrt{2}}(\ket{2_a0_b} + e^{2i\phi}\ket{0_a2_b}),
    \label{eqn:N00N}
\end{equation}
where the phase $\phi$ can be set with an on-chip thermo-optic phase shifter.
The state $\ket{\psi}$ requires the two SPDC sources have equal photon-pair generation probability, and that the generated photons are indistinguishable.
After setting the phase $\phi$, the path-entangled state passes through a Mach-Zehnder interferometer (MZI) consisting of two directional couplers ($H'$) and a phase shifter ($R$).
A MZI acts as a tunable beamsplitter with the operation $U_{MZI} = H'\cdot R(\theta)\cdot H'$, where 
\begin{equation}
    H' = \frac{1}{\sqrt{2}}\begin{bmatrix}
        1 & i\\
        i & 1
    \end{bmatrix},\,\,R(\theta) = \begin{bmatrix}
        1 & 0\\
        0 & e^{i\theta}
    \end{bmatrix},
\end{equation}
and the phase $\theta$ is controlled by a second thermo-optic phase shifter.
With $\theta=\tfrac{(2n+1)\pi}{2}$ for $n\in\mathbb{N}$, the MZI behaves as a 50:50 beamsplitter and the chip generates the state
\begin{align}
    \ket{\psi(\theta=\tfrac{(2n+1)\pi}{2})} = & \cos{(\phi)}\tfrac{1}{\sqrt{2}}(\ket{2_a0_b} - \ket{0_a2_b}) \nonumber \\
    + & \sin{(\phi)}\ket{1_a1_b}.
    \label{eqn:psi_out}
\end{align}
This indicates that the photons will bunch with $\phi=m\pi$ for $m\in\mathbb{N}$ and will anti-bunch with $\phi=\tfrac{(2m+1)\pi}{2}$, exiting the device at separate output ports.
The impact of an unbalanced $N00N$ state and reduced photon indistinguishability are explored in Supplementary Section 1.

We fabricated our device using an x-cut LNOI wafer stack with a \SI{300}{nm} lithium niobate thin film, \SI{4.7}{\um} thermally grown silicon dioxide and a \SI{525}{\um} silicon handle.
The cross-section of the waveguide and thermo-optic phase shifter are shown in Figure \ref{fig:schematic}b.
We simulated the optical modes, shown in Figure \ref{fig:schematic}c, using a finite element mode solver to calculate the required \SI{2.478}{\um} poling period for type-0 quasi phase matching.
We first deposited comb-shaped chromium electrodes on the surface of the LNOI wafer and applied voltage pulses to locally invert the lithium niobate crystal axis.
The poling region is \SI{2}{mm} long and the electrodes are separated by \SI{30}{\um}, which is sufficient to accommodate two isolated waveguides.
By increasing the separation of the poling electrodes, we could reasonably fit around 10 waveguides in a single poling region without cross coupling \cite{wang_ultrahigh-efficiency_2018}.
More details on the poling process are given in Supplementary Section 2.
After poling, we fabricated waveguides by electron-beam lithography (EBL) of a HSQ-based mask and inductively coupled plasma reactive ion etching in argon, which leads to the typical trapezoidal waveguides observed in LNOI photonics \cite{kaufmann_redeposition-free_2023}.
Lastly, we deposited gold thermo-optic phase shifters via another EBL step followed by metal evaporation and lift-off.

\begin{figure}
    \centering
    \includegraphics[width=1\linewidth]{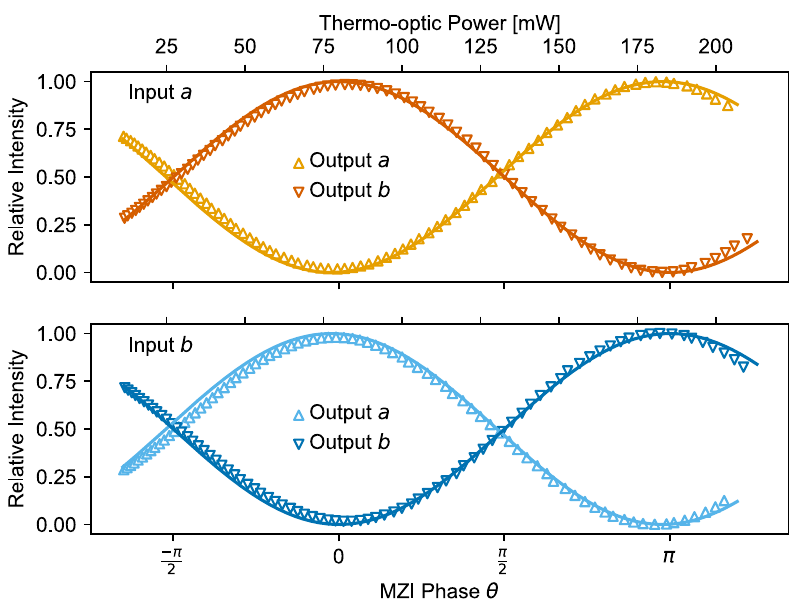}
    \caption{\textbf{MZI classical characterization.} Laser light at \SI{1562}{nm} is injected into one input mode of the MZI at a time. The phase $\theta$ is tuned and we monitor the optical power at both outputs of the MZI.}
    \label{fig:classical}
\end{figure}

\begin{figure}
    \centering
    \includegraphics[width=\linewidth]{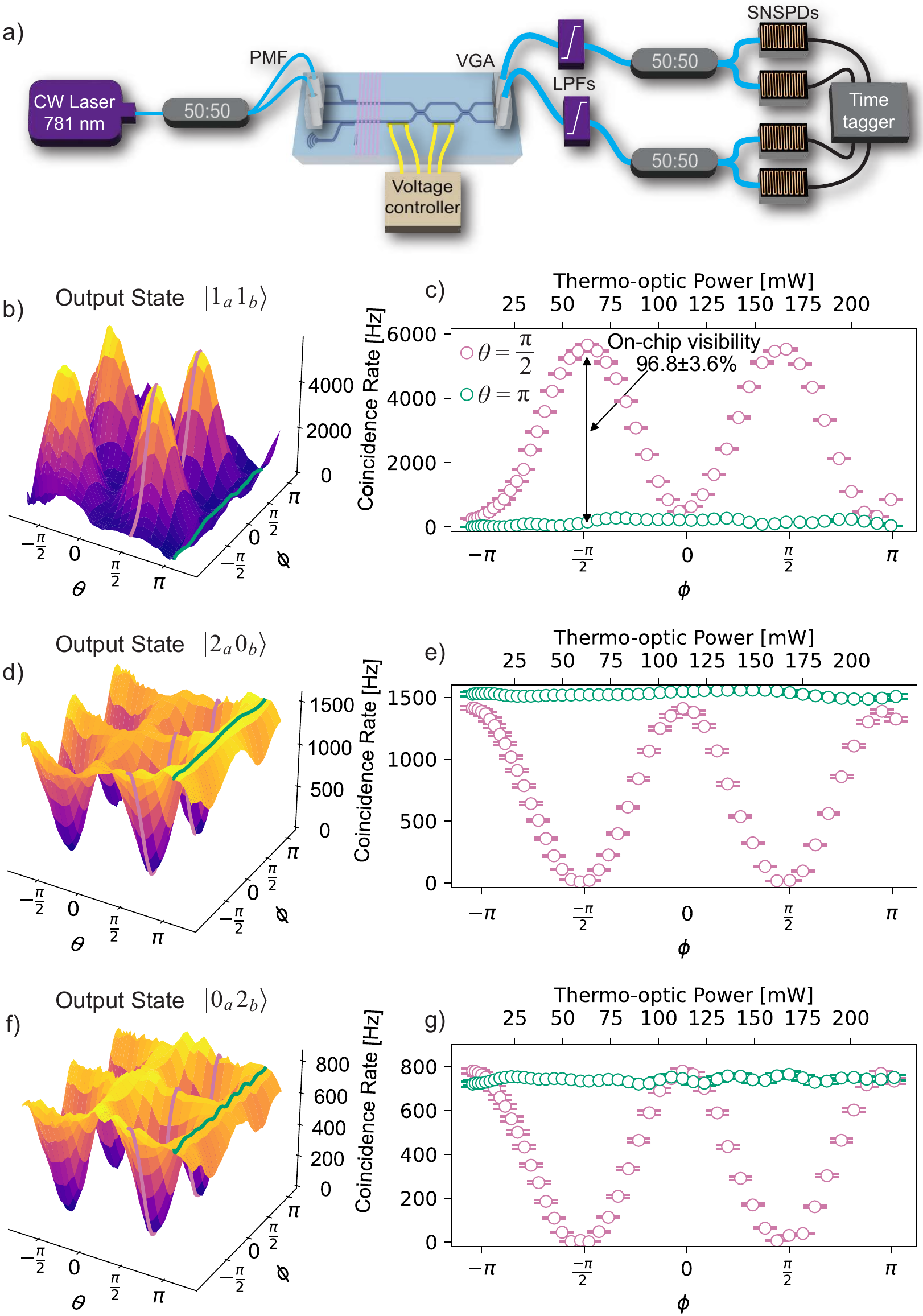}
    \caption{\textbf{On-chip quantum interference.} 
    a) Schematic of the experiment.
    We pump both SPDC sources with the same laser using a 50:50 fiber beamsplitter. 
    We control each phase with a programmable voltage source.
    At the output we use two more 50:50 fiber beamsplitters to measure all three output basis states $\{\ket{2_a0_b},\ket{1_a1_b},\ket{0_a2_b}\}$. PMF: Polarization maintaining fiber, VGA: V-groove fiber array, LPF: Long-pass filter, SNSPD: Superconducting nanowire single photon detector.
    b), c) Output probability for state $\ket{1_a1_b}$ as a function of the two phases $\theta$ and $\phi$.
    d), e) Output probability for state $\ket{2_a0_b}$ as a function of the two phases $\theta$ and $\phi$.
    f), g) Output probability for state $\ket{0_a2_b}$ as a function of the two phases $\theta$ and $\phi$.
    With $\theta=\tfrac{(2n+1)\pi}{2}$ we observe strong bunching and anti-bunching, depending on the phase of the $N00N$ state $\phi$.
    With $\theta=n\pi$ we directly observe the bunched $N00N$ state generated by SPDC.
    }
    \label{fig:quantum}
\end{figure}

We classically characterize the device to measure the phase matching of the two periodically poled waveguides and the tuning of the MZI.
To characterize the phase-matching of the periodically poled waveguides, we couple a tunable telecom wavelength laser backwards through the chip and measure the up-converted second-harmonic light collected at the input grating couplers, as shown in Figures \ref{fig:schematic}d and \ref{fig:schematic}e.
Both waveguides are within the same periodically poled region of the LNOI chip and we therefore expect near-identical phase matching.
Indeed both waveguides have the maximum second-harmonic generation at $\sim\SI{1562}{nm}$ and we attribute the mismatch in intensity to alignment of the fiber to the grating coupler rather than differences between the waveguides.
Even with a mismatch in nonlinear efficiency, we can compensate the pump power in each waveguide to generate a balanced $N00N$ state.
The second lobe observed in the up-converted signal is likely caused by variations in the lithium niobate film thickness over the length of the periodically poled region, which can be overcome by adapting the poling period along the length of the waveguide \cite{chen_adapted_2024}.
The bandwidth of the second-harmonic generation is related to the length of the periodically poled waveguide, however, the bandwidth of the generated photons is expected to be much broader due to the phase matching relation in type-0 SPDC, which could lead to reduced visibility of the on-chip interferometer \cite{graffitti_design_2018}.
We classically characterize the on-chip MZI by coupling laser at \SI{1562}{nm} wavelength to the chip using the on-chip wavelength division multiplexers, such that we can use grating couplers at the design wavelength.
We scan the voltage of the phase shifter $\theta$ using a programmable voltage source, and measure the applied electrical power and optical output in waveguide $a$ and $b$ simultaneously using a v-groove fiber array and multi-channel power meter.
In Figure \ref{fig:classical} we plot the measured optical signal for both inputs and both outputs, and calculate \SI{98.2(17)}{\percent} average visibility across all measurements, which suggests the directional couplers have near-ideal 50:50 splitting ratio.

\begin{figure}
    \centering
    \includegraphics[width=1\linewidth]{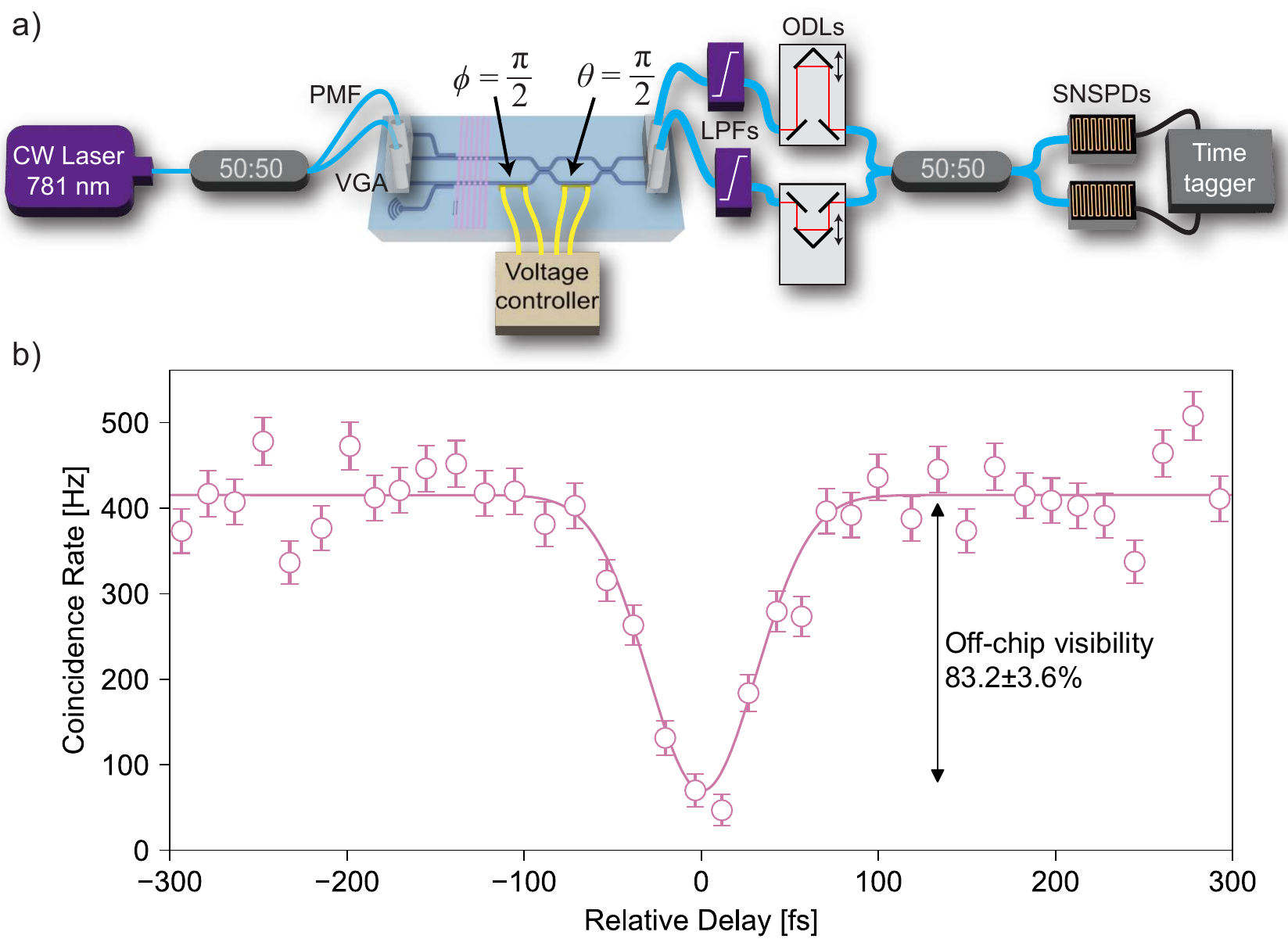}
    \caption{\textbf{Hong-Ou-Mandel interference} a) Experimental setup for measuring Hong-Ou-Mandel interference. b) Coincidence rate with $\theta=\phi=\tfrac{\pi}{2}$ while scanning optical delay lines (ODLs) with the characteristic Hong-Ou-Mandel dip at zero relative delay.}
    \label{fig:hom}
\end{figure}

To generate the two-photon $N00N$ state, we inject a continuous-wave laser at \SI{781}{nm} into both periodically poled waveguides simultaneously, as shown in Figure \ref{fig:quantum}a.
We use long pass filters after the chip to reject the pump, two fiber splitters and four superconducting single photon detectors to record the three two-photon states $\ket{1_a1_b}$, $\ket{2_a0_b}$ and $\ket{0_a2_b}$.
We scan the two on-chip phase shifters $\theta$ and $\phi$, and record the coincidence rates for each state with a time tagger.
In Figure \ref{fig:quantum}b we plot the $\ket{1_a1_b}$ and observe peaks that correspond to anti-bunching when $\theta=\tfrac{(2n+1)\pi}{2}$ and $\phi=\tfrac{(2m+1)\pi}{2}$.
These peaks correspond to minima in the $\ket{2_a0_b}$ and $\ket{0_a2_b}$ states, as shown in Figures \ref{fig:quantum}d and \ref{fig:quantum}f respectively.
Conversely, when $\theta=n\pi$ or $\phi=m\pi$ we observe bunching at the output of the device.
To highlight the visibility of the interference we plot the three output states as a function of $\phi$ for $\theta=\tfrac{\pi}{2}$ and $\theta=\pi$ in Figures \ref{fig:quantum}c, \ref{fig:quantum}e and \ref{fig:quantum}g respectively.
We calculate an average visibility across all three states of \SI{96.8(3.6)}{\percent}, suggesting that we are generating a balanced $N00N$ state, as given in Equation \ref{eqn:N00N}, the photons are indistinguishable, and the linear optic circuit performs well despite the broad bandwidth expected from the SPDC source.
We measure very high photon-pair generation rate with our periodically poled LNOI waveguides.
We pump the SPDC process with an estimated $\sim\SI{10}{\uW}$ on-chip laser power and measure coincidence rates $>\SI{5800}{pairs/second}$ at the detectors.
Given the output losses of $\sim\SI{13}{dB}$ from the grating couplers (\SI{10}{dB}), long-pass filters (\SI{1}{dB}), fiber beamsplitters (\SI{1}{dB}) and single photon detectors (\SI{1}{dB}), we estimate an on-chip brightness of $\sim\SI{2.3e8}{pairs/s/mW}$.
We neglect the typical waveguide loss of $\sim\SI{0.2}{dB/cm}$ as it adds very little to the overall loss of the device.
This value is significantly higher than the leading free-space SPDC sources based on ppKTP crystals, even if we consider the broad bandwidth of our photon-pairs \cite{jabir_robust_2017, meraner_approaching_2021}, and could achieve even higher values \cite{xue_ultrabright_2021}.
The out-coupling efficiency can be improved by reducing the grating coupler losses \cite{krasnokutska_high_2019}, for example by including a back reflector \cite{chen_low-loss_2022}.
We expect the maximum signal for $\ket{2_a0_b}$ and $\ket{0_a2_b}$ to be four times lower than the $\ket{1_a1_b}$ due of the nature of the state and the probabilistic separation in the fiber beamsplitters.
However, we record a lower coincidence rate for the $\ket{0_a2_b}$ than expected, which we attribute to misalignment between the grating and the fiber.


We finally use an off-chip HOM experiment to verify the generation of indistinguishable photons in the two output modes of the chip.
To this end, we optically pump both waveguides and set the two phase shifters to $\theta=\phi=\tfrac{\pi}{2}$ such that the output state is $\ket{1_a1_b}$.
The photons then pass through two tunable optical delay lines before interfering in a fiber beamsplitter.
We measure the degree of photon indistinguishability by tuning the relative optical delay and recording the characteristic Hong-Ou-Mandel dip, as shown in Figure \ref{fig:hom}.
With zero time-delay, the photons bunch into one output with a visibility of \SI{83.2(1)}{\percent}.
We measure the width of the HOM dip as \SI{71.9(0.5)}{fs}, which suggests a spectral bandwidth of the photons, after coupling into fiber, of $\sim\SI{50}{\nm}$. 
The visibility is reduced in comparison to the on-chip quantum interference measured in Figure \ref{fig:quantum}, which we attribute to the limited bandwidth of the fiber beamsplitter.

\section*{Conclusion}
Integrated photonic circuits provide a route towards scalable quantum information processing for tasks such as Gaussian boson sampling which has application in graph searches \cite{yu_universal_2023}, vibronic spectra simulations \cite{huh_boson_2015}, and drug discovery \cite{sempere-llagostera_experimentally_2022}.
While significant progress has been made with silicon and silicon nitride photonics for quantum computation \cite{adcock_advances_2021, arrazola_quantum_2021}, these platforms have drawbacks such as a weak nonlinearity and two-photon absorption that may ultimately limit their applicability \cite{silverstone_silicon_2016}.
We have demonstrated that lithium niobate-on-insulator can also be applied to quantum information tasks by generating a two-photon path-entangled state and performing on-chip quantum interference in a programmable interferometer.
Including high-speed electro-optic switching could enable a quantum photonics platform capable of SPDC source multiplexing and feedforward operations for quantum error correction \cite{saravi_lithium_2021}.
It is now feasible to scale up our technology to many sources and large waveguide circuits for near-term quantum photonics experiments such as cluster state generation, boson sampling, and quantum enhanced metrology.

\subsection*{Data availability}

Raw data and evaluation code are available from the authors upon reasonable request.

\subsection*{Competing interests}
The authors declare no competing financial or non-financial interests.

\subsection*{Author Contributions}

R.J.C. conceived the experiment and designed the chip. T.K., J.K., A.S., A.M., G.F. and F.K fabricated the lithium niobate-on-insulator sample including periodic poling, lithography, etching and metal deposition. R.J.C. performed the optical measurements, data analysis and wrote the original draft of the manuscript. R.G. supervised the project. All authors contributed to revising the manuscript.

\subsection*{Acknowledgments}

We acknowledge support for characterization of our samples from the Scientific Center of Optical and Electron Microscopy ScopeM and from the cleanroom facilities BRNC and FIRST of ETH Zurich and IBM Ruschlikon.
R.J.C. acknowledges support from the Swiss National Science Foundation under the Ambizione Fellowship Program (Project Number 208707).
R.G. acknowledges support from the European Space Agency (Project Number 4000137426), the Swiss National Science Foundation under the Bridge Program (Project Number 194693), and the European Research Council (Project Number 714837).


\begin{thebibliography}{37}%
	\makeatletter
	\providecommand \@ifxundefined [1]{%
		\@ifx{#1\undefined}
	}%
	\providecommand \@ifnum [1]{%
		\ifnum #1\expandafter \@firstoftwo
		\else \expandafter \@secondoftwo
		\fi
	}%
	\providecommand \@ifx [1]{%
		\ifx #1\expandafter \@firstoftwo
		\else \expandafter \@secondoftwo
		\fi
	}%
	\providecommand \natexlab [1]{#1}%
	\providecommand \enquote  [1]{``#1''}%
	\providecommand \bibnamefont  [1]{#1}%
	\providecommand \bibfnamefont [1]{#1}%
	\providecommand \citenamefont [1]{#1}%
	\providecommand \href@noop [0]{\@secondoftwo}%
	\providecommand \href [0]{\begingroup \@sanitize@url \@href}%
	\providecommand \@href[1]{\@@startlink{#1}\@@href}%
	\providecommand \@@href[1]{\endgroup#1\@@endlink}%
	\providecommand \@sanitize@url [0]{\catcode `\\12\catcode `\$12\catcode `\&12\catcode `\#12\catcode `\^12\catcode `\_12\catcode `\%12\relax}%
	\providecommand \@@startlink[1]{}%
	\providecommand \@@endlink[0]{}%
	\providecommand \url  [0]{\begingroup\@sanitize@url \@url }%
	\providecommand \@url [1]{\endgroup\@href {#1}{\urlprefix }}%
	\providecommand \urlprefix  [0]{URL }%
	\providecommand \Eprint [0]{\href }%
	\providecommand \doibase [0]{https://doi.org/}%
	\providecommand \selectlanguage [0]{\@gobble}%
	\providecommand \bibinfo  [0]{\@secondoftwo}%
	\providecommand \bibfield  [0]{\@secondoftwo}%
	\providecommand \translation [1]{[#1]}%
	\providecommand \BibitemOpen [0]{}%
	\providecommand \bibitemStop [0]{}%
	\providecommand \bibitemNoStop [0]{.\EOS\space}%
	\providecommand \EOS [0]{\spacefactor3000\relax}%
	\providecommand \BibitemShut  [1]{\csname bibitem#1\endcsname}%
	\let\auto@bib@innerbib\@empty
	\bibitem [{\citenamefont {Zhong}\ \emph {et~al.}(2020)\citenamefont {Zhong}, \citenamefont {Wang}, \citenamefont {Deng}, \citenamefont {Chen}, \citenamefont {Peng}, \citenamefont {Luo}, \citenamefont {Qin}, \citenamefont {Wu}, \citenamefont {Ding}, \citenamefont {Hu}, \citenamefont {Hu}, \citenamefont {Yang}, \citenamefont {Zhang}, \citenamefont {Li}, \citenamefont {Li}, \citenamefont {Jiang}, \citenamefont {Gan}, \citenamefont {Yang}, \citenamefont {You}, \citenamefont {Wang}, \citenamefont {Li}, \citenamefont {Liu}, \citenamefont {Lu},\ and\ \citenamefont {Pan}}]{zhong_quantum_2020}%
	\BibitemOpen
	\bibfield  {author} {\bibinfo {author} {\bibfnamefont {H.-S.}\ \bibnamefont {Zhong}}, \bibinfo {author} {\bibfnamefont {H.}~\bibnamefont {Wang}}, \bibinfo {author} {\bibfnamefont {Y.-H.}\ \bibnamefont {Deng}}, \bibinfo {author} {\bibfnamefont {M.-C.}\ \bibnamefont {Chen}}, \bibinfo {author} {\bibfnamefont {L.-C.}\ \bibnamefont {Peng}}, \bibinfo {author} {\bibfnamefont {Y.-H.}\ \bibnamefont {Luo}}, \bibinfo {author} {\bibfnamefont {J.}~\bibnamefont {Qin}}, \bibinfo {author} {\bibfnamefont {D.}~\bibnamefont {Wu}}, \bibinfo {author} {\bibfnamefont {X.}~\bibnamefont {Ding}}, \bibinfo {author} {\bibfnamefont {Y.}~\bibnamefont {Hu}}, \bibinfo {author} {\bibfnamefont {P.}~\bibnamefont {Hu}}, \bibinfo {author} {\bibfnamefont {X.-Y.}\ \bibnamefont {Yang}}, \bibinfo {author} {\bibfnamefont {W.-J.}\ \bibnamefont {Zhang}}, \bibinfo {author} {\bibfnamefont {H.}~\bibnamefont {Li}}, \bibinfo {author} {\bibfnamefont {Y.}~\bibnamefont {Li}}, \bibinfo {author} {\bibfnamefont {X.}~\bibnamefont {Jiang}}, \bibinfo {author}
		{\bibfnamefont {L.}~\bibnamefont {Gan}}, \bibinfo {author} {\bibfnamefont {G.}~\bibnamefont {Yang}}, \bibinfo {author} {\bibfnamefont {L.}~\bibnamefont {You}}, \bibinfo {author} {\bibfnamefont {Z.}~\bibnamefont {Wang}}, \bibinfo {author} {\bibfnamefont {L.}~\bibnamefont {Li}}, \bibinfo {author} {\bibfnamefont {N.-L.}\ \bibnamefont {Liu}}, \bibinfo {author} {\bibfnamefont {C.-Y.}\ \bibnamefont {Lu}},\ and\ \bibinfo {author} {\bibfnamefont {J.-W.}\ \bibnamefont {Pan}},\ }\bibfield  {title} {\bibinfo {title} {Quantum computational advantage using photons},\ }\href {https://doi.org/10.1126/science.abe8770} {\bibfield  {journal} {\bibinfo  {journal} {Science}\ }\textbf {\bibinfo {volume} {370}},\ \bibinfo {pages} {1460} (\bibinfo {year} {2020})}\BibitemShut {NoStop}%
	\bibitem [{\citenamefont {Madsen}\ \emph {et~al.}(2022)\citenamefont {Madsen}, \citenamefont {Laudenbach}, \citenamefont {Askarani}, \citenamefont {Rortais}, \citenamefont {Vincent}, \citenamefont {Bulmer}, \citenamefont {Miatto}, \citenamefont {Neuhaus}, \citenamefont {Helt}, \citenamefont {Collins}, \citenamefont {Lita}, \citenamefont {Gerrits}, \citenamefont {Nam}, \citenamefont {Vaidya}, \citenamefont {Menotti}, \citenamefont {Dhand}, \citenamefont {Vernon}, \citenamefont {Quesada},\ and\ \citenamefont {Lavoie}}]{madsen_quantum_2022}%
	\BibitemOpen
	\bibfield  {author} {\bibinfo {author} {\bibfnamefont {L.~S.}\ \bibnamefont {Madsen}}, \bibinfo {author} {\bibfnamefont {F.}~\bibnamefont {Laudenbach}}, \bibinfo {author} {\bibfnamefont {M.~F.}\ \bibnamefont {Askarani}}, \bibinfo {author} {\bibfnamefont {F.}~\bibnamefont {Rortais}}, \bibinfo {author} {\bibfnamefont {T.}~\bibnamefont {Vincent}}, \bibinfo {author} {\bibfnamefont {J.~F.~F.}\ \bibnamefont {Bulmer}}, \bibinfo {author} {\bibfnamefont {F.~M.}\ \bibnamefont {Miatto}}, \bibinfo {author} {\bibfnamefont {L.}~\bibnamefont {Neuhaus}}, \bibinfo {author} {\bibfnamefont {L.~G.}\ \bibnamefont {Helt}}, \bibinfo {author} {\bibfnamefont {M.~J.}\ \bibnamefont {Collins}}, \bibinfo {author} {\bibfnamefont {A.~E.}\ \bibnamefont {Lita}}, \bibinfo {author} {\bibfnamefont {T.}~\bibnamefont {Gerrits}}, \bibinfo {author} {\bibfnamefont {S.~W.}\ \bibnamefont {Nam}}, \bibinfo {author} {\bibfnamefont {V.~D.}\ \bibnamefont {Vaidya}}, \bibinfo {author} {\bibfnamefont {M.}~\bibnamefont {Menotti}}, \bibinfo {author}
		{\bibfnamefont {I.}~\bibnamefont {Dhand}}, \bibinfo {author} {\bibfnamefont {Z.}~\bibnamefont {Vernon}}, \bibinfo {author} {\bibfnamefont {N.}~\bibnamefont {Quesada}},\ and\ \bibinfo {author} {\bibfnamefont {J.}~\bibnamefont {Lavoie}},\ }\bibfield  {title} {\bibinfo {title} {Quantum computational advantage with a programmable photonic processor},\ }\href {https://doi.org/10.1038/s41586-022-04725-x} {\bibfield  {journal} {\bibinfo  {journal} {Nature}\ }\textbf {\bibinfo {volume} {606}},\ \bibinfo {pages} {75} (\bibinfo {year} {2022})}\BibitemShut {NoStop}%
	\bibitem [{\citenamefont {Deng}\ \emph {et~al.}(2023)\citenamefont {Deng}, \citenamefont {Gu}, \citenamefont {Liu}, \citenamefont {Gong}, \citenamefont {Su}, \citenamefont {Zhang}, \citenamefont {Tang}, \citenamefont {Jia}, \citenamefont {Xu}, \citenamefont {Chen}, \citenamefont {Qin}, \citenamefont {Peng}, \citenamefont {Yan}, \citenamefont {Hu}, \citenamefont {Huang}, \citenamefont {Li}, \citenamefont {Li}, \citenamefont {Chen}, \citenamefont {Jiang}, \citenamefont {Gan}, \citenamefont {Yang}, \citenamefont {You}, \citenamefont {Li}, \citenamefont {Zhong}, \citenamefont {Wang}, \citenamefont {Liu}, \citenamefont {Renema}, \citenamefont {Lu},\ and\ \citenamefont {Pan}}]{deng_gaussian_2023}%
	\BibitemOpen
	\bibfield  {author} {\bibinfo {author} {\bibfnamefont {Y.-H.}\ \bibnamefont {Deng}}, \bibinfo {author} {\bibfnamefont {Y.-C.}\ \bibnamefont {Gu}}, \bibinfo {author} {\bibfnamefont {H.-L.}\ \bibnamefont {Liu}}, \bibinfo {author} {\bibfnamefont {S.-Q.}\ \bibnamefont {Gong}}, \bibinfo {author} {\bibfnamefont {H.}~\bibnamefont {Su}}, \bibinfo {author} {\bibfnamefont {Z.-J.}\ \bibnamefont {Zhang}}, \bibinfo {author} {\bibfnamefont {H.-Y.}\ \bibnamefont {Tang}}, \bibinfo {author} {\bibfnamefont {M.-H.}\ \bibnamefont {Jia}}, \bibinfo {author} {\bibfnamefont {J.-M.}\ \bibnamefont {Xu}}, \bibinfo {author} {\bibfnamefont {M.-C.}\ \bibnamefont {Chen}}, \bibinfo {author} {\bibfnamefont {J.}~\bibnamefont {Qin}}, \bibinfo {author} {\bibfnamefont {L.-C.}\ \bibnamefont {Peng}}, \bibinfo {author} {\bibfnamefont {J.}~\bibnamefont {Yan}}, \bibinfo {author} {\bibfnamefont {Y.}~\bibnamefont {Hu}}, \bibinfo {author} {\bibfnamefont {J.}~\bibnamefont {Huang}}, \bibinfo {author} {\bibfnamefont {H.}~\bibnamefont {Li}}, \bibinfo
		{author} {\bibfnamefont {Y.}~\bibnamefont {Li}}, \bibinfo {author} {\bibfnamefont {Y.}~\bibnamefont {Chen}}, \bibinfo {author} {\bibfnamefont {X.}~\bibnamefont {Jiang}}, \bibinfo {author} {\bibfnamefont {L.}~\bibnamefont {Gan}}, \bibinfo {author} {\bibfnamefont {G.}~\bibnamefont {Yang}}, \bibinfo {author} {\bibfnamefont {L.}~\bibnamefont {You}}, \bibinfo {author} {\bibfnamefont {L.}~\bibnamefont {Li}}, \bibinfo {author} {\bibfnamefont {H.-S.}\ \bibnamefont {Zhong}}, \bibinfo {author} {\bibfnamefont {H.}~\bibnamefont {Wang}}, \bibinfo {author} {\bibfnamefont {N.-L.}\ \bibnamefont {Liu}}, \bibinfo {author} {\bibfnamefont {J.~J.}\ \bibnamefont {Renema}}, \bibinfo {author} {\bibfnamefont {C.-Y.}\ \bibnamefont {Lu}},\ and\ \bibinfo {author} {\bibfnamefont {J.-W.}\ \bibnamefont {Pan}},\ }\bibfield  {title} {\bibinfo {title} {Gaussian {Boson} {Sampling} with {Pseudo}-{Photon}-{Number}-{Resolving} {Detectors} and {Quantum} {Computational} {Advantage}},\ }\href {https://doi.org/10.1103/PhysRevLett.131.150601}
	{\bibfield  {journal} {\bibinfo  {journal} {Physical Review Letters}\ }\textbf {\bibinfo {volume} {131}},\ \bibinfo {pages} {150601} (\bibinfo {year} {2023})}\BibitemShut {NoStop}%
	\bibitem [{\citenamefont {Wang}\ \emph {et~al.}(2020)\citenamefont {Wang}, \citenamefont {Sciarrino}, \citenamefont {Laing},\ and\ \citenamefont {Thompson}}]{wang_integrated_2020}%
	\BibitemOpen
	\bibfield  {author} {\bibinfo {author} {\bibfnamefont {J.}~\bibnamefont {Wang}}, \bibinfo {author} {\bibfnamefont {F.}~\bibnamefont {Sciarrino}}, \bibinfo {author} {\bibfnamefont {A.}~\bibnamefont {Laing}},\ and\ \bibinfo {author} {\bibfnamefont {M.~G.}\ \bibnamefont {Thompson}},\ }\bibfield  {title} {\bibinfo {title} {Integrated photonic quantum technologies},\ }\href {https://doi.org/10.1038/s41566-019-0532-1} {\bibfield  {journal} {\bibinfo  {journal} {Nature Photonics}\ }\textbf {\bibinfo {volume} {14}},\ \bibinfo {pages} {273} (\bibinfo {year} {2020})}\BibitemShut {NoStop}%
	\bibitem [{\citenamefont {Adcock}\ \emph {et~al.}(2021)\citenamefont {Adcock}, \citenamefont {Bao}, \citenamefont {Chi}, \citenamefont {Chen}, \citenamefont {Bacco}, \citenamefont {Gong}, \citenamefont {Oxenløwe}, \citenamefont {Wang},\ and\ \citenamefont {Ding}}]{adcock_advances_2021}%
	\BibitemOpen
	\bibfield  {author} {\bibinfo {author} {\bibfnamefont {J.~C.}\ \bibnamefont {Adcock}}, \bibinfo {author} {\bibfnamefont {J.}~\bibnamefont {Bao}}, \bibinfo {author} {\bibfnamefont {Y.}~\bibnamefont {Chi}}, \bibinfo {author} {\bibfnamefont {X.}~\bibnamefont {Chen}}, \bibinfo {author} {\bibfnamefont {D.}~\bibnamefont {Bacco}}, \bibinfo {author} {\bibfnamefont {Q.}~\bibnamefont {Gong}}, \bibinfo {author} {\bibfnamefont {L.~K.}\ \bibnamefont {Oxenløwe}}, \bibinfo {author} {\bibfnamefont {J.}~\bibnamefont {Wang}},\ and\ \bibinfo {author} {\bibfnamefont {Y.}~\bibnamefont {Ding}},\ }\bibfield  {title} {\bibinfo {title} {Advances in {Silicon} {Quantum} {Photonics}},\ }\href {https://doi.org/10.1109/JSTQE.2020.3025737} {\bibfield  {journal} {\bibinfo  {journal} {IEEE Journal of Selected Topics in Quantum Electronics}\ }\textbf {\bibinfo {volume} {27}},\ \bibinfo {pages} {1} (\bibinfo {year} {2021})}\BibitemShut {NoStop}%
	\bibitem [{\citenamefont {Arrazola}\ \emph {et~al.}(2021)\citenamefont {Arrazola}, \citenamefont {Bergholm}, \citenamefont {Brádler}, \citenamefont {Bromley}, \citenamefont {Collins}, \citenamefont {Dhand}, \citenamefont {Fumagalli}, \citenamefont {Gerrits}, \citenamefont {Goussev}, \citenamefont {Helt}, \citenamefont {Hundal}, \citenamefont {Isacsson}, \citenamefont {Israel}, \citenamefont {Izaac}, \citenamefont {Jahangiri}, \citenamefont {Janik}, \citenamefont {Killoran}, \citenamefont {Kumar}, \citenamefont {Lavoie}, \citenamefont {Lita}, \citenamefont {Mahler}, \citenamefont {Menotti}, \citenamefont {Morrison}, \citenamefont {Nam}, \citenamefont {Neuhaus}, \citenamefont {Qi}, \citenamefont {Quesada}, \citenamefont {Repingon}, \citenamefont {Sabapathy}, \citenamefont {Schuld}, \citenamefont {Su}, \citenamefont {Swinarton}, \citenamefont {Száva}, \citenamefont {Tan}, \citenamefont {Tan}, \citenamefont {Vaidya}, \citenamefont {Vernon}, \citenamefont {Zabaneh},\ and\ \citenamefont
		{Zhang}}]{arrazola_quantum_2021}%
	\BibitemOpen
	\bibfield  {author} {\bibinfo {author} {\bibfnamefont {J.~M.}\ \bibnamefont {Arrazola}}, \bibinfo {author} {\bibfnamefont {V.}~\bibnamefont {Bergholm}}, \bibinfo {author} {\bibfnamefont {K.}~\bibnamefont {Brádler}}, \bibinfo {author} {\bibfnamefont {T.~R.}\ \bibnamefont {Bromley}}, \bibinfo {author} {\bibfnamefont {M.~J.}\ \bibnamefont {Collins}}, \bibinfo {author} {\bibfnamefont {I.}~\bibnamefont {Dhand}}, \bibinfo {author} {\bibfnamefont {A.}~\bibnamefont {Fumagalli}}, \bibinfo {author} {\bibfnamefont {T.}~\bibnamefont {Gerrits}}, \bibinfo {author} {\bibfnamefont {A.}~\bibnamefont {Goussev}}, \bibinfo {author} {\bibfnamefont {L.~G.}\ \bibnamefont {Helt}}, \bibinfo {author} {\bibfnamefont {J.}~\bibnamefont {Hundal}}, \bibinfo {author} {\bibfnamefont {T.}~\bibnamefont {Isacsson}}, \bibinfo {author} {\bibfnamefont {R.~B.}\ \bibnamefont {Israel}}, \bibinfo {author} {\bibfnamefont {J.}~\bibnamefont {Izaac}}, \bibinfo {author} {\bibfnamefont {S.}~\bibnamefont {Jahangiri}}, \bibinfo {author} {\bibfnamefont
			{R.}~\bibnamefont {Janik}}, \bibinfo {author} {\bibfnamefont {N.}~\bibnamefont {Killoran}}, \bibinfo {author} {\bibfnamefont {S.~P.}\ \bibnamefont {Kumar}}, \bibinfo {author} {\bibfnamefont {J.}~\bibnamefont {Lavoie}}, \bibinfo {author} {\bibfnamefont {A.~E.}\ \bibnamefont {Lita}}, \bibinfo {author} {\bibfnamefont {D.~H.}\ \bibnamefont {Mahler}}, \bibinfo {author} {\bibfnamefont {M.}~\bibnamefont {Menotti}}, \bibinfo {author} {\bibfnamefont {B.}~\bibnamefont {Morrison}}, \bibinfo {author} {\bibfnamefont {S.~W.}\ \bibnamefont {Nam}}, \bibinfo {author} {\bibfnamefont {L.}~\bibnamefont {Neuhaus}}, \bibinfo {author} {\bibfnamefont {H.~Y.}\ \bibnamefont {Qi}}, \bibinfo {author} {\bibfnamefont {N.}~\bibnamefont {Quesada}}, \bibinfo {author} {\bibfnamefont {A.}~\bibnamefont {Repingon}}, \bibinfo {author} {\bibfnamefont {K.~K.}\ \bibnamefont {Sabapathy}}, \bibinfo {author} {\bibfnamefont {M.}~\bibnamefont {Schuld}}, \bibinfo {author} {\bibfnamefont {D.}~\bibnamefont {Su}}, \bibinfo {author} {\bibfnamefont
			{J.}~\bibnamefont {Swinarton}}, \bibinfo {author} {\bibfnamefont {A.}~\bibnamefont {Száva}}, \bibinfo {author} {\bibfnamefont {K.}~\bibnamefont {Tan}}, \bibinfo {author} {\bibfnamefont {P.}~\bibnamefont {Tan}}, \bibinfo {author} {\bibfnamefont {V.~D.}\ \bibnamefont {Vaidya}}, \bibinfo {author} {\bibfnamefont {Z.}~\bibnamefont {Vernon}}, \bibinfo {author} {\bibfnamefont {Z.}~\bibnamefont {Zabaneh}},\ and\ \bibinfo {author} {\bibfnamefont {Y.}~\bibnamefont {Zhang}},\ }\bibfield  {title} {\bibinfo {title} {Quantum circuits with many photons on a programmable nanophotonic chip},\ }\href {https://doi.org/10.1038/s41586-021-03202-1} {\bibfield  {journal} {\bibinfo  {journal} {Nature}\ }\textbf {\bibinfo {volume} {591}},\ \bibinfo {pages} {54} (\bibinfo {year} {2021})}\BibitemShut {NoStop}%
	\bibitem [{\citenamefont {Bao}\ \emph {et~al.}(2023)\citenamefont {Bao}, \citenamefont {Fu}, \citenamefont {Pramanik}, \citenamefont {Mao}, \citenamefont {Chi}, \citenamefont {Cao}, \citenamefont {Zhai}, \citenamefont {Mao}, \citenamefont {Dai}, \citenamefont {Chen}, \citenamefont {Jia}, \citenamefont {Zhao}, \citenamefont {Zheng}, \citenamefont {Tang}, \citenamefont {Li}, \citenamefont {Luo}, \citenamefont {Wang}, \citenamefont {Yang}, \citenamefont {Peng}, \citenamefont {Liu}, \citenamefont {Dai}, \citenamefont {He}, \citenamefont {Muthali}, \citenamefont {Oxenløwe}, \citenamefont {Vigliar}, \citenamefont {Paesani}, \citenamefont {Hou}, \citenamefont {Santagati}, \citenamefont {Silverstone}, \citenamefont {Laing}, \citenamefont {Thompson}, \citenamefont {O’Brien}, \citenamefont {Ding}, \citenamefont {Gong},\ and\ \citenamefont {Wang}}]{bao_very-large-scale_2023}%
	\BibitemOpen
	\bibfield  {author} {\bibinfo {author} {\bibfnamefont {J.}~\bibnamefont {Bao}}, \bibinfo {author} {\bibfnamefont {Z.}~\bibnamefont {Fu}}, \bibinfo {author} {\bibfnamefont {T.}~\bibnamefont {Pramanik}}, \bibinfo {author} {\bibfnamefont {J.}~\bibnamefont {Mao}}, \bibinfo {author} {\bibfnamefont {Y.}~\bibnamefont {Chi}}, \bibinfo {author} {\bibfnamefont {Y.}~\bibnamefont {Cao}}, \bibinfo {author} {\bibfnamefont {C.}~\bibnamefont {Zhai}}, \bibinfo {author} {\bibfnamefont {Y.}~\bibnamefont {Mao}}, \bibinfo {author} {\bibfnamefont {T.}~\bibnamefont {Dai}}, \bibinfo {author} {\bibfnamefont {X.}~\bibnamefont {Chen}}, \bibinfo {author} {\bibfnamefont {X.}~\bibnamefont {Jia}}, \bibinfo {author} {\bibfnamefont {L.}~\bibnamefont {Zhao}}, \bibinfo {author} {\bibfnamefont {Y.}~\bibnamefont {Zheng}}, \bibinfo {author} {\bibfnamefont {B.}~\bibnamefont {Tang}}, \bibinfo {author} {\bibfnamefont {Z.}~\bibnamefont {Li}}, \bibinfo {author} {\bibfnamefont {J.}~\bibnamefont {Luo}}, \bibinfo {author} {\bibfnamefont
			{W.}~\bibnamefont {Wang}}, \bibinfo {author} {\bibfnamefont {Y.}~\bibnamefont {Yang}}, \bibinfo {author} {\bibfnamefont {Y.}~\bibnamefont {Peng}}, \bibinfo {author} {\bibfnamefont {D.}~\bibnamefont {Liu}}, \bibinfo {author} {\bibfnamefont {D.}~\bibnamefont {Dai}}, \bibinfo {author} {\bibfnamefont {Q.}~\bibnamefont {He}}, \bibinfo {author} {\bibfnamefont {A.~L.}\ \bibnamefont {Muthali}}, \bibinfo {author} {\bibfnamefont {L.~K.}\ \bibnamefont {Oxenløwe}}, \bibinfo {author} {\bibfnamefont {C.}~\bibnamefont {Vigliar}}, \bibinfo {author} {\bibfnamefont {S.}~\bibnamefont {Paesani}}, \bibinfo {author} {\bibfnamefont {H.}~\bibnamefont {Hou}}, \bibinfo {author} {\bibfnamefont {R.}~\bibnamefont {Santagati}}, \bibinfo {author} {\bibfnamefont {J.~W.}\ \bibnamefont {Silverstone}}, \bibinfo {author} {\bibfnamefont {A.}~\bibnamefont {Laing}}, \bibinfo {author} {\bibfnamefont {M.~G.}\ \bibnamefont {Thompson}}, \bibinfo {author} {\bibfnamefont {J.~L.}\ \bibnamefont {O’Brien}}, \bibinfo {author} {\bibfnamefont
			{Y.}~\bibnamefont {Ding}}, \bibinfo {author} {\bibfnamefont {Q.}~\bibnamefont {Gong}},\ and\ \bibinfo {author} {\bibfnamefont {J.}~\bibnamefont {Wang}},\ }\bibfield  {title} {\bibinfo {title} {Very-large-scale integrated quantum graph photonics},\ }\href {https://doi.org/10.1038/s41566-023-01187-z} {\bibfield  {journal} {\bibinfo  {journal} {Nature Photonics}\ ,\ \bibinfo {pages} {1}} (\bibinfo {year} {2023})}\BibitemShut {NoStop}%
	\bibitem [{\citenamefont {Zhong}\ \emph {et~al.}(2018)\citenamefont {Zhong}, \citenamefont {Li}, \citenamefont {Li}, \citenamefont {Peng}, \citenamefont {Su}, \citenamefont {Hu}, \citenamefont {He}, \citenamefont {Ding}, \citenamefont {Zhang}, \citenamefont {Li}, \citenamefont {Zhang}, \citenamefont {Wang}, \citenamefont {You}, \citenamefont {Wang}, \citenamefont {Jiang}, \citenamefont {Li}, \citenamefont {Chen}, \citenamefont {Liu}, \citenamefont {Lu},\ and\ \citenamefont {Pan}}]{zhong_12-photon_2018}%
	\BibitemOpen
	\bibfield  {author} {\bibinfo {author} {\bibfnamefont {H.-S.}\ \bibnamefont {Zhong}}, \bibinfo {author} {\bibfnamefont {Y.}~\bibnamefont {Li}}, \bibinfo {author} {\bibfnamefont {W.}~\bibnamefont {Li}}, \bibinfo {author} {\bibfnamefont {L.-C.}\ \bibnamefont {Peng}}, \bibinfo {author} {\bibfnamefont {Z.-E.}\ \bibnamefont {Su}}, \bibinfo {author} {\bibfnamefont {Y.}~\bibnamefont {Hu}}, \bibinfo {author} {\bibfnamefont {Y.-M.}\ \bibnamefont {He}}, \bibinfo {author} {\bibfnamefont {X.}~\bibnamefont {Ding}}, \bibinfo {author} {\bibfnamefont {W.}~\bibnamefont {Zhang}}, \bibinfo {author} {\bibfnamefont {H.}~\bibnamefont {Li}}, \bibinfo {author} {\bibfnamefont {L.}~\bibnamefont {Zhang}}, \bibinfo {author} {\bibfnamefont {Z.}~\bibnamefont {Wang}}, \bibinfo {author} {\bibfnamefont {L.}~\bibnamefont {You}}, \bibinfo {author} {\bibfnamefont {X.-L.}\ \bibnamefont {Wang}}, \bibinfo {author} {\bibfnamefont {X.}~\bibnamefont {Jiang}}, \bibinfo {author} {\bibfnamefont {L.}~\bibnamefont {Li}}, \bibinfo {author} {\bibfnamefont
			{Y.-A.}\ \bibnamefont {Chen}}, \bibinfo {author} {\bibfnamefont {N.-L.}\ \bibnamefont {Liu}}, \bibinfo {author} {\bibfnamefont {C.-Y.}\ \bibnamefont {Lu}},\ and\ \bibinfo {author} {\bibfnamefont {J.-W.}\ \bibnamefont {Pan}},\ }\bibfield  {title} {\bibinfo {title} {12-{Photon} {Entanglement} and {Scalable} {Scattershot} {Boson} {Sampling} with {Optimal} {Entangled}-{Photon} {Pairs} from {Parametric} {Down}-{Conversion}},\ }\href {https://doi.org/10.1103/PhysRevLett.121.250505} {\bibfield  {journal} {\bibinfo  {journal} {Physical Review Letters}\ }\textbf {\bibinfo {volume} {121}},\ \bibinfo {pages} {250505} (\bibinfo {year} {2018})}\BibitemShut {NoStop}%
	\bibitem [{\citenamefont {Konno}\ \emph {et~al.}(2024)\citenamefont {Konno}, \citenamefont {Asavanant}, \citenamefont {Hanamura}, \citenamefont {Nagayoshi}, \citenamefont {Fukui}, \citenamefont {Sakaguchi}, \citenamefont {Ide}, \citenamefont {China}, \citenamefont {Yabuno}, \citenamefont {Miki}, \citenamefont {Terai}, \citenamefont {Takase}, \citenamefont {Endo}, \citenamefont {Marek}, \citenamefont {Filip}, \citenamefont {van Loock},\ and\ \citenamefont {Furusawa}}]{konno_logical_2024}%
	\BibitemOpen
	\bibfield  {author} {\bibinfo {author} {\bibfnamefont {S.}~\bibnamefont {Konno}}, \bibinfo {author} {\bibfnamefont {W.}~\bibnamefont {Asavanant}}, \bibinfo {author} {\bibfnamefont {F.}~\bibnamefont {Hanamura}}, \bibinfo {author} {\bibfnamefont {H.}~\bibnamefont {Nagayoshi}}, \bibinfo {author} {\bibfnamefont {K.}~\bibnamefont {Fukui}}, \bibinfo {author} {\bibfnamefont {A.}~\bibnamefont {Sakaguchi}}, \bibinfo {author} {\bibfnamefont {R.}~\bibnamefont {Ide}}, \bibinfo {author} {\bibfnamefont {F.}~\bibnamefont {China}}, \bibinfo {author} {\bibfnamefont {M.}~\bibnamefont {Yabuno}}, \bibinfo {author} {\bibfnamefont {S.}~\bibnamefont {Miki}}, \bibinfo {author} {\bibfnamefont {H.}~\bibnamefont {Terai}}, \bibinfo {author} {\bibfnamefont {K.}~\bibnamefont {Takase}}, \bibinfo {author} {\bibfnamefont {M.}~\bibnamefont {Endo}}, \bibinfo {author} {\bibfnamefont {P.}~\bibnamefont {Marek}}, \bibinfo {author} {\bibfnamefont {R.}~\bibnamefont {Filip}}, \bibinfo {author} {\bibfnamefont {P.}~\bibnamefont {van Loock}},\ and\
		\bibinfo {author} {\bibfnamefont {A.}~\bibnamefont {Furusawa}},\ }\bibfield  {title} {\bibinfo {title} {Logical states for fault-tolerant quantum computation with propagating light},\ }\href {https://doi.org/10.1126/science.adk7560} {\bibfield  {journal} {\bibinfo  {journal} {Science}\ }\textbf {\bibinfo {volume} {383}},\ \bibinfo {pages} {289} (\bibinfo {year} {2024})}\BibitemShut {NoStop}%
	\bibitem [{\citenamefont {Zhu}\ \emph {et~al.}(2021)\citenamefont {Zhu}, \citenamefont {Shao}, \citenamefont {Yu}, \citenamefont {Cheng}, \citenamefont {Desiatov}, \citenamefont {Xin}, \citenamefont {Hu}, \citenamefont {Holzgrafe}, \citenamefont {Ghosh}, \citenamefont {Shams-Ansari}, \citenamefont {Puma}, \citenamefont {Sinclair}, \citenamefont {Reimer}, \citenamefont {Zhang},\ and\ \citenamefont {Lončar}}]{zhu_integrated_2021}%
	\BibitemOpen
	\bibfield  {author} {\bibinfo {author} {\bibfnamefont {D.}~\bibnamefont {Zhu}}, \bibinfo {author} {\bibfnamefont {L.}~\bibnamefont {Shao}}, \bibinfo {author} {\bibfnamefont {M.}~\bibnamefont {Yu}}, \bibinfo {author} {\bibfnamefont {R.}~\bibnamefont {Cheng}}, \bibinfo {author} {\bibfnamefont {B.}~\bibnamefont {Desiatov}}, \bibinfo {author} {\bibfnamefont {C.~J.}\ \bibnamefont {Xin}}, \bibinfo {author} {\bibfnamefont {Y.}~\bibnamefont {Hu}}, \bibinfo {author} {\bibfnamefont {J.}~\bibnamefont {Holzgrafe}}, \bibinfo {author} {\bibfnamefont {S.}~\bibnamefont {Ghosh}}, \bibinfo {author} {\bibfnamefont {A.}~\bibnamefont {Shams-Ansari}}, \bibinfo {author} {\bibfnamefont {E.}~\bibnamefont {Puma}}, \bibinfo {author} {\bibfnamefont {N.}~\bibnamefont {Sinclair}}, \bibinfo {author} {\bibfnamefont {C.}~\bibnamefont {Reimer}}, \bibinfo {author} {\bibfnamefont {M.}~\bibnamefont {Zhang}},\ and\ \bibinfo {author} {\bibfnamefont {M.}~\bibnamefont {Lončar}},\ }\bibfield  {title} {\bibinfo {title} {Integrated photonics on
			thin-film lithium niobate},\ }\href {https://doi.org/10.1364/AOP.411024} {\bibfield  {journal} {\bibinfo  {journal} {Advances in Optics and Photonics}\ }\textbf {\bibinfo {volume} {13}},\ \bibinfo {pages} {242} (\bibinfo {year} {2021})}\BibitemShut {NoStop}%
	\bibitem [{\citenamefont {Wang}\ \emph {et~al.}(2018{\natexlab{a}})\citenamefont {Wang}, \citenamefont {Zhang}, \citenamefont {Chen}, \citenamefont {Bertrand}, \citenamefont {Shams-Ansari}, \citenamefont {Chandrasekhar}, \citenamefont {Winzer},\ and\ \citenamefont {Lončar}}]{wang_integrated_2018}%
	\BibitemOpen
	\bibfield  {author} {\bibinfo {author} {\bibfnamefont {C.}~\bibnamefont {Wang}}, \bibinfo {author} {\bibfnamefont {M.}~\bibnamefont {Zhang}}, \bibinfo {author} {\bibfnamefont {X.}~\bibnamefont {Chen}}, \bibinfo {author} {\bibfnamefont {M.}~\bibnamefont {Bertrand}}, \bibinfo {author} {\bibfnamefont {A.}~\bibnamefont {Shams-Ansari}}, \bibinfo {author} {\bibfnamefont {S.}~\bibnamefont {Chandrasekhar}}, \bibinfo {author} {\bibfnamefont {P.}~\bibnamefont {Winzer}},\ and\ \bibinfo {author} {\bibfnamefont {M.}~\bibnamefont {Lončar}},\ }\bibfield  {title} {\bibinfo {title} {Integrated lithium niobate electro-optic modulators operating at {CMOS}-compatible voltages},\ }\href {https://doi.org/10.1038/s41586-018-0551-y} {\bibfield  {journal} {\bibinfo  {journal} {Nature}\ ,\ \bibinfo {pages} {1}} (\bibinfo {year} {2018}{\natexlab{a}})}\BibitemShut {NoStop}%
	\bibitem [{\citenamefont {Wang}\ \emph {et~al.}(2018{\natexlab{b}})\citenamefont {Wang}, \citenamefont {Langrock}, \citenamefont {Marandi}, \citenamefont {Jankowski}, \citenamefont {Zhang}, \citenamefont {Desiatov}, \citenamefont {Fejer},\ and\ \citenamefont {Lončar}}]{wang_ultrahigh-efficiency_2018}%
	\BibitemOpen
	\bibfield  {author} {\bibinfo {author} {\bibfnamefont {C.}~\bibnamefont {Wang}}, \bibinfo {author} {\bibfnamefont {C.}~\bibnamefont {Langrock}}, \bibinfo {author} {\bibfnamefont {A.}~\bibnamefont {Marandi}}, \bibinfo {author} {\bibfnamefont {M.}~\bibnamefont {Jankowski}}, \bibinfo {author} {\bibfnamefont {M.}~\bibnamefont {Zhang}}, \bibinfo {author} {\bibfnamefont {B.}~\bibnamefont {Desiatov}}, \bibinfo {author} {\bibfnamefont {M.~M.}\ \bibnamefont {Fejer}},\ and\ \bibinfo {author} {\bibfnamefont {M.}~\bibnamefont {Lončar}},\ }\bibfield  {title} {\bibinfo {title} {Ultrahigh-efficiency wavelength conversion in nanophotonic periodically poled lithium niobate waveguides},\ }\href {https://doi.org/10.1364/OPTICA.5.001438} {\bibfield  {journal} {\bibinfo  {journal} {Optica}\ }\textbf {\bibinfo {volume} {5}},\ \bibinfo {pages} {1438} (\bibinfo {year} {2018}{\natexlab{b}})}\BibitemShut {NoStop}%
	\bibitem [{\citenamefont {Lu}\ \emph {et~al.}(2019)\citenamefont {Lu}, \citenamefont {Surya}, \citenamefont {Liu}, \citenamefont {Bruch}, \citenamefont {Gong}, \citenamefont {Xu},\ and\ \citenamefont {Tang}}]{lu_periodically_2019}%
	\BibitemOpen
	\bibfield  {author} {\bibinfo {author} {\bibfnamefont {J.}~\bibnamefont {Lu}}, \bibinfo {author} {\bibfnamefont {J.~B.}\ \bibnamefont {Surya}}, \bibinfo {author} {\bibfnamefont {X.}~\bibnamefont {Liu}}, \bibinfo {author} {\bibfnamefont {A.~W.}\ \bibnamefont {Bruch}}, \bibinfo {author} {\bibfnamefont {Z.}~\bibnamefont {Gong}}, \bibinfo {author} {\bibfnamefont {Y.}~\bibnamefont {Xu}},\ and\ \bibinfo {author} {\bibfnamefont {H.~X.}\ \bibnamefont {Tang}},\ }\bibfield  {title} {\bibinfo {title} {Periodically poled thin-film lithium niobate microring resonators with a second-harmonic generation efficiency of 250,000\%/{W}},\ }\href {https://doi.org/10.1364/OPTICA.6.001455} {\bibfield  {journal} {\bibinfo  {journal} {Optica}\ }\textbf {\bibinfo {volume} {6}},\ \bibinfo {pages} {1455} (\bibinfo {year} {2019})}\BibitemShut {NoStop}%
	\bibitem [{\citenamefont {Zhao}\ \emph {et~al.}(2020)\citenamefont {Zhao}, \citenamefont {Ma}, \citenamefont {Rüsing},\ and\ \citenamefont {Mookherjea}}]{zhao_high_2020}%
	\BibitemOpen
	\bibfield  {author} {\bibinfo {author} {\bibfnamefont {J.}~\bibnamefont {Zhao}}, \bibinfo {author} {\bibfnamefont {C.}~\bibnamefont {Ma}}, \bibinfo {author} {\bibfnamefont {M.}~\bibnamefont {Rüsing}},\ and\ \bibinfo {author} {\bibfnamefont {S.}~\bibnamefont {Mookherjea}},\ }\bibfield  {title} {\bibinfo {title} {High {Quality} {Entangled} {Photon} {Pair} {Generation} in {Periodically} {Poled} {Thin}-{Film} {Lithium} {Niobate} {Waveguides}},\ }\href {https://doi.org/10.1103/PhysRevLett.124.163603} {\bibfield  {journal} {\bibinfo  {journal} {Physical Review Letters}\ }\textbf {\bibinfo {volume} {124}},\ \bibinfo {pages} {163603} (\bibinfo {year} {2020})}\BibitemShut {NoStop}%
	\bibitem [{\citenamefont {Xin}\ \emph {et~al.}(2022)\citenamefont {Xin}, \citenamefont {Xin}, \citenamefont {Mishra}, \citenamefont {Mishra}, \citenamefont {Chen}, \citenamefont {Zhu}, \citenamefont {Shams-Ansari}, \citenamefont {Langrock}, \citenamefont {Sinclair}, \citenamefont {Sinclair}, \citenamefont {Wong}, \citenamefont {Fejer},\ and\ \citenamefont {Lončar}}]{xin_spectrally_2022}%
	\BibitemOpen
	\bibfield  {author} {\bibinfo {author} {\bibfnamefont {C.~J.}\ \bibnamefont {Xin}}, \bibinfo {author} {\bibfnamefont {C.~J.}\ \bibnamefont {Xin}}, \bibinfo {author} {\bibfnamefont {J.}~\bibnamefont {Mishra}}, \bibinfo {author} {\bibfnamefont {J.}~\bibnamefont {Mishra}}, \bibinfo {author} {\bibfnamefont {C.}~\bibnamefont {Chen}}, \bibinfo {author} {\bibfnamefont {D.}~\bibnamefont {Zhu}}, \bibinfo {author} {\bibfnamefont {A.}~\bibnamefont {Shams-Ansari}}, \bibinfo {author} {\bibfnamefont {C.}~\bibnamefont {Langrock}}, \bibinfo {author} {\bibfnamefont {N.}~\bibnamefont {Sinclair}}, \bibinfo {author} {\bibfnamefont {N.}~\bibnamefont {Sinclair}}, \bibinfo {author} {\bibfnamefont {F.~N.~C.}\ \bibnamefont {Wong}}, \bibinfo {author} {\bibfnamefont {M.~M.}\ \bibnamefont {Fejer}},\ and\ \bibinfo {author} {\bibfnamefont {M.}~\bibnamefont {Lončar}},\ }\bibfield  {title} {\bibinfo {title} {Spectrally separable photon-pair generation in dispersion engineered thin-film lithium niobate},\ }\href
	{https://doi.org/10.1364/OL.456873} {\bibfield  {journal} {\bibinfo  {journal} {Optics Letters}\ }\textbf {\bibinfo {volume} {47}},\ \bibinfo {pages} {2830} (\bibinfo {year} {2022})}\BibitemShut {NoStop}%
	\bibitem [{\citenamefont {Chapman}\ \emph {et~al.}(2023)\citenamefont {Chapman}, \citenamefont {Häusler}, \citenamefont {Finco}, \citenamefont {Kaufmann},\ and\ \citenamefont {Grange}}]{chapman_quantum_2023}%
	\BibitemOpen
	\bibfield  {author} {\bibinfo {author} {\bibfnamefont {R.~J.}\ \bibnamefont {Chapman}}, \bibinfo {author} {\bibfnamefont {S.}~\bibnamefont {Häusler}}, \bibinfo {author} {\bibfnamefont {G.}~\bibnamefont {Finco}}, \bibinfo {author} {\bibfnamefont {F.}~\bibnamefont {Kaufmann}},\ and\ \bibinfo {author} {\bibfnamefont {R.}~\bibnamefont {Grange}},\ }\bibfield  {title} {\bibinfo {title} {Quantum logical controlled-{NOT} gate in a lithium niobate-on-insulator photonic quantum walk},\ }\href {https://doi.org/10.1088/2058-9565/ad0a48} {\bibfield  {journal} {\bibinfo  {journal} {Quantum Science and Technology}\ }\textbf {\bibinfo {volume} {9}},\ \bibinfo {pages} {015016} (\bibinfo {year} {2023})}\BibitemShut {NoStop}%
	\bibitem [{\citenamefont {Babel}\ \emph {et~al.}(2023)\citenamefont {Babel}, \citenamefont {Bollmers}, \citenamefont {Massaro}, \citenamefont {Luo}, \citenamefont {Stefszky}, \citenamefont {Pegoraro}, \citenamefont {Held}, \citenamefont {Herrmann}, \citenamefont {Eigner}, \citenamefont {Brecht}, \citenamefont {Padberg},\ and\ \citenamefont {Silberhorn}}]{babel_demonstration_2023}%
	\BibitemOpen
	\bibfield  {author} {\bibinfo {author} {\bibfnamefont {S.}~\bibnamefont {Babel}}, \bibinfo {author} {\bibfnamefont {L.}~\bibnamefont {Bollmers}}, \bibinfo {author} {\bibfnamefont {M.}~\bibnamefont {Massaro}}, \bibinfo {author} {\bibfnamefont {K.~H.}\ \bibnamefont {Luo}}, \bibinfo {author} {\bibfnamefont {M.}~\bibnamefont {Stefszky}}, \bibinfo {author} {\bibfnamefont {F.}~\bibnamefont {Pegoraro}}, \bibinfo {author} {\bibfnamefont {P.}~\bibnamefont {Held}}, \bibinfo {author} {\bibfnamefont {H.}~\bibnamefont {Herrmann}}, \bibinfo {author} {\bibfnamefont {C.}~\bibnamefont {Eigner}}, \bibinfo {author} {\bibfnamefont {B.}~\bibnamefont {Brecht}}, \bibinfo {author} {\bibfnamefont {L.}~\bibnamefont {Padberg}},\ and\ \bibinfo {author} {\bibfnamefont {C.}~\bibnamefont {Silberhorn}},\ }\bibfield  {title} {\bibinfo {title} {Demonstration of {Hong}-{Ou}-{Mandel} interference in an {LNOI} directional coupler},\ }\href {https://doi.org/10.1364/OE.484126} {\bibfield  {journal} {\bibinfo  {journal} {Optics Express}\ }\textbf
		{\bibinfo {volume} {31}},\ \bibinfo {pages} {23140} (\bibinfo {year} {2023})}\BibitemShut {NoStop}%
	\bibitem [{\citenamefont {Silverstone}\ \emph {et~al.}(2014)\citenamefont {Silverstone}, \citenamefont {Bonneau}, \citenamefont {Ohira}, \citenamefont {Suzuki}, \citenamefont {Yoshida}, \citenamefont {Iizuka}, \citenamefont {Ezaki}, \citenamefont {Natarajan}, \citenamefont {Tanner}, \citenamefont {Hadfield}, \citenamefont {Zwiller}, \citenamefont {Marshall}, \citenamefont {Rarity}, \citenamefont {O'Brien},\ and\ \citenamefont {Thompson}}]{silverstone_-chip_2014}%
	\BibitemOpen
	\bibfield  {author} {\bibinfo {author} {\bibfnamefont {J.~W.}\ \bibnamefont {Silverstone}}, \bibinfo {author} {\bibfnamefont {D.}~\bibnamefont {Bonneau}}, \bibinfo {author} {\bibfnamefont {K.}~\bibnamefont {Ohira}}, \bibinfo {author} {\bibfnamefont {N.}~\bibnamefont {Suzuki}}, \bibinfo {author} {\bibfnamefont {H.}~\bibnamefont {Yoshida}}, \bibinfo {author} {\bibfnamefont {N.}~\bibnamefont {Iizuka}}, \bibinfo {author} {\bibfnamefont {M.}~\bibnamefont {Ezaki}}, \bibinfo {author} {\bibfnamefont {C.~M.}\ \bibnamefont {Natarajan}}, \bibinfo {author} {\bibfnamefont {M.~G.}\ \bibnamefont {Tanner}}, \bibinfo {author} {\bibfnamefont {R.~H.}\ \bibnamefont {Hadfield}}, \bibinfo {author} {\bibfnamefont {V.}~\bibnamefont {Zwiller}}, \bibinfo {author} {\bibfnamefont {G.~D.}\ \bibnamefont {Marshall}}, \bibinfo {author} {\bibfnamefont {J.~G.}\ \bibnamefont {Rarity}}, \bibinfo {author} {\bibfnamefont {J.~L.}\ \bibnamefont {O'Brien}},\ and\ \bibinfo {author} {\bibfnamefont {M.~G.}\ \bibnamefont {Thompson}},\ }\bibfield
	{title} {\bibinfo {title} {On-chip quantum interference between silicon photon-pair sources},\ }\href {https://doi.org/10.1038/nphoton.2013.339} {\bibfield  {journal} {\bibinfo  {journal} {Nature Photonics}\ }\textbf {\bibinfo {volume} {8}},\ \bibinfo {pages} {104} (\bibinfo {year} {2014})}\BibitemShut {NoStop}%
	\bibitem [{\citenamefont {Knill}\ \emph {et~al.}(2001)\citenamefont {Knill}, \citenamefont {Laflamme},\ and\ \citenamefont {Milburn}}]{knill_scheme_2001}%
	\BibitemOpen
	\bibfield  {author} {\bibinfo {author} {\bibfnamefont {E.}~\bibnamefont {Knill}}, \bibinfo {author} {\bibfnamefont {R.}~\bibnamefont {Laflamme}},\ and\ \bibinfo {author} {\bibfnamefont {G.~J.}\ \bibnamefont {Milburn}},\ }\bibfield  {title} {\bibinfo {title} {A scheme for efficient quantum computation with linear optics},\ }\href {https://doi.org/10.1038/35051009} {\bibfield  {journal} {\bibinfo  {journal} {Nature}\ }\textbf {\bibinfo {volume} {409}},\ \bibinfo {pages} {46} (\bibinfo {year} {2001})}\BibitemShut {NoStop}%
	\bibitem [{\citenamefont {Ralph}\ \emph {et~al.}(2002)\citenamefont {Ralph}, \citenamefont {Langford}, \citenamefont {Bell},\ and\ \citenamefont {White}}]{ralph_linear_2002}%
	\BibitemOpen
	\bibfield  {author} {\bibinfo {author} {\bibfnamefont {T.~C.}\ \bibnamefont {Ralph}}, \bibinfo {author} {\bibfnamefont {N.~K.}\ \bibnamefont {Langford}}, \bibinfo {author} {\bibfnamefont {T.~B.}\ \bibnamefont {Bell}},\ and\ \bibinfo {author} {\bibfnamefont {A.~G.}\ \bibnamefont {White}},\ }\bibfield  {title} {\bibinfo {title} {Linear optical controlled-{NOT} gate in the coincidence basis},\ }\href {https://doi.org/10.1103/PhysRevA.65.062324} {\bibfield  {journal} {\bibinfo  {journal} {Physical Review A}\ }\textbf {\bibinfo {volume} {65}},\ \bibinfo {pages} {062324} (\bibinfo {year} {2002})}\BibitemShut {NoStop}%
	\bibitem [{\citenamefont {Taballione}\ \emph {et~al.}(2023)\citenamefont {Taballione}, \citenamefont {Anguita}, \citenamefont {Goede}, \citenamefont {Venderbosch}, \citenamefont {Kassenberg}, \citenamefont {Snijders}, \citenamefont {Kannan}, \citenamefont {Vleeshouwers}, \citenamefont {Smith}, \citenamefont {Epping}, \citenamefont {Meer}, \citenamefont {Pinkse}, \citenamefont {Vlekkert},\ and\ \citenamefont {Renema}}]{taballione_20-mode_2023}%
	\BibitemOpen
	\bibfield  {author} {\bibinfo {author} {\bibfnamefont {C.}~\bibnamefont {Taballione}}, \bibinfo {author} {\bibfnamefont {M.~C.}\ \bibnamefont {Anguita}}, \bibinfo {author} {\bibfnamefont {M.~d.}\ \bibnamefont {Goede}}, \bibinfo {author} {\bibfnamefont {P.}~\bibnamefont {Venderbosch}}, \bibinfo {author} {\bibfnamefont {B.}~\bibnamefont {Kassenberg}}, \bibinfo {author} {\bibfnamefont {H.}~\bibnamefont {Snijders}}, \bibinfo {author} {\bibfnamefont {N.}~\bibnamefont {Kannan}}, \bibinfo {author} {\bibfnamefont {W.~L.}\ \bibnamefont {Vleeshouwers}}, \bibinfo {author} {\bibfnamefont {D.}~\bibnamefont {Smith}}, \bibinfo {author} {\bibfnamefont {J.~P.}\ \bibnamefont {Epping}}, \bibinfo {author} {\bibfnamefont {R.~v.~d.}\ \bibnamefont {Meer}}, \bibinfo {author} {\bibfnamefont {P.~W.~H.}\ \bibnamefont {Pinkse}}, \bibinfo {author} {\bibfnamefont {H.~v.~d.}\ \bibnamefont {Vlekkert}},\ and\ \bibinfo {author} {\bibfnamefont {J.~J.}\ \bibnamefont {Renema}},\ }\bibfield  {title} {\bibinfo {title} {20-{Mode} {Universal}
			{Quantum} {Photonic} {Processor}},\ }\href {https://doi.org/10.22331/q-2023-08-01-1071} {\bibfield  {journal} {\bibinfo  {journal} {Quantum}\ }\textbf {\bibinfo {volume} {7}},\ \bibinfo {pages} {1071} (\bibinfo {year} {2023})}\BibitemShut {NoStop}%
	\bibitem [{\citenamefont {Hong}\ \emph {et~al.}(1987)\citenamefont {Hong}, \citenamefont {Ou},\ and\ \citenamefont {Mandel}}]{hong_measurement_1987}%
	\BibitemOpen
	\bibfield  {author} {\bibinfo {author} {\bibfnamefont {C.~K.}\ \bibnamefont {Hong}}, \bibinfo {author} {\bibfnamefont {Z.~Y.}\ \bibnamefont {Ou}},\ and\ \bibinfo {author} {\bibfnamefont {L.}~\bibnamefont {Mandel}},\ }\bibfield  {title} {\bibinfo {title} {Measurement of subpicosecond time intervals between two photons by interference},\ }\href {https://doi.org/10.1103/PhysRevLett.59.2044} {\bibfield  {journal} {\bibinfo  {journal} {Physical Review Letters}\ }\textbf {\bibinfo {volume} {59}},\ \bibinfo {pages} {2044} (\bibinfo {year} {1987})}\BibitemShut {NoStop}%
	\bibitem [{\citenamefont {Dowling}(2008)}]{dowling_quantum_2008}%
	\BibitemOpen
	\bibfield  {author} {\bibinfo {author} {\bibfnamefont {J.~P.}\ \bibnamefont {Dowling}},\ }\bibfield  {title} {\bibinfo {title} {Quantum optical metrology – the lowdown on high-{N00N} states},\ }\href {https://doi.org/10.1080/00107510802091298} {\bibfield  {journal} {\bibinfo  {journal} {Contemporary Physics}\ }\textbf {\bibinfo {volume} {49}},\ \bibinfo {pages} {125} (\bibinfo {year} {2008})}\BibitemShut {NoStop}%
	\bibitem [{\citenamefont {Jin}\ \emph {et~al.}(2014)\citenamefont {Jin}, \citenamefont {Liu}, \citenamefont {Xu}, \citenamefont {Xia}, \citenamefont {Zhong}, \citenamefont {Yuan}, \citenamefont {Zhou}, \citenamefont {Gong}, \citenamefont {Wang},\ and\ \citenamefont {Zhu}}]{jin_-chip_2014}%
	\BibitemOpen
	\bibfield  {author} {\bibinfo {author} {\bibfnamefont {H.}~\bibnamefont {Jin}}, \bibinfo {author} {\bibfnamefont {F.}~\bibnamefont {Liu}}, \bibinfo {author} {\bibfnamefont {P.}~\bibnamefont {Xu}}, \bibinfo {author} {\bibfnamefont {J.}~\bibnamefont {Xia}}, \bibinfo {author} {\bibfnamefont {M.}~\bibnamefont {Zhong}}, \bibinfo {author} {\bibfnamefont {Y.}~\bibnamefont {Yuan}}, \bibinfo {author} {\bibfnamefont {J.}~\bibnamefont {Zhou}}, \bibinfo {author} {\bibfnamefont {Y.}~\bibnamefont {Gong}}, \bibinfo {author} {\bibfnamefont {W.}~\bibnamefont {Wang}},\ and\ \bibinfo {author} {\bibfnamefont {S.}~\bibnamefont {Zhu}},\ }\bibfield  {title} {\bibinfo {title} {On-{Chip} {Generation} and {Manipulation} of {Entangled} {Photons} {Based} on {Reconfigurable} {Lithium}-{Niobate} {Waveguide} {Circuits}},\ }\href {https://doi.org/10.1103/PhysRevLett.113.103601} {\bibfield  {journal} {\bibinfo  {journal} {Physical Review Letters}\ }\textbf {\bibinfo {volume} {113}},\ \bibinfo {pages} {103601} (\bibinfo {year}
		{2014})}\BibitemShut {NoStop}%
	\bibitem [{\citenamefont {Kaufmann}\ \emph {et~al.}(2023)\citenamefont {Kaufmann}, \citenamefont {Finco}, \citenamefont {Maeder},\ and\ \citenamefont {Grange}}]{kaufmann_redeposition-free_2023}%
	\BibitemOpen
	\bibfield  {author} {\bibinfo {author} {\bibfnamefont {F.}~\bibnamefont {Kaufmann}}, \bibinfo {author} {\bibfnamefont {G.}~\bibnamefont {Finco}}, \bibinfo {author} {\bibfnamefont {A.}~\bibnamefont {Maeder}},\ and\ \bibinfo {author} {\bibfnamefont {R.}~\bibnamefont {Grange}},\ }\bibfield  {title} {\bibinfo {title} {Redeposition-free inductively-coupled plasma etching of lithium niobate for integrated photonics},\ }\bibfield  {journal} {\bibinfo  {journal} {Nanophotonics}\ }\href {https://doi.org/10.1515/nanoph-2022-0676} {10.1515/nanoph-2022-0676} (\bibinfo {year} {2023})\BibitemShut {NoStop}%
	\bibitem [{\citenamefont {Chen}\ \emph {et~al.}(2024)\citenamefont {Chen}, \citenamefont {Briggs}, \citenamefont {Cui}, \citenamefont {Zhang}, \citenamefont {Shah},\ and\ \citenamefont {Fan}}]{chen_adapted_2024}%
	\BibitemOpen
	\bibfield  {author} {\bibinfo {author} {\bibfnamefont {P.-K.}\ \bibnamefont {Chen}}, \bibinfo {author} {\bibfnamefont {I.}~\bibnamefont {Briggs}}, \bibinfo {author} {\bibfnamefont {C.}~\bibnamefont {Cui}}, \bibinfo {author} {\bibfnamefont {L.}~\bibnamefont {Zhang}}, \bibinfo {author} {\bibfnamefont {M.}~\bibnamefont {Shah}},\ and\ \bibinfo {author} {\bibfnamefont {L.}~\bibnamefont {Fan}},\ }\bibfield  {title} {\bibinfo {title} {Adapted poling to break the nonlinear efficiency limit in nanophotonic lithium niobate waveguides},\ }\href {https://doi.org/10.1038/s41565-023-01525-w} {\bibfield  {journal} {\bibinfo  {journal} {Nature Nanotechnology}\ }\textbf {\bibinfo {volume} {19}},\ \bibinfo {pages} {44} (\bibinfo {year} {2024})}\BibitemShut {NoStop}%
	\bibitem [{\citenamefont {Graffitti}\ \emph {et~al.}(2018)\citenamefont {Graffitti}, \citenamefont {Kelly-Massicotte}, \citenamefont {Fedrizzi},\ and\ \citenamefont {Brańczyk}}]{graffitti_design_2018}%
	\BibitemOpen
	\bibfield  {author} {\bibinfo {author} {\bibfnamefont {F.}~\bibnamefont {Graffitti}}, \bibinfo {author} {\bibfnamefont {J.}~\bibnamefont {Kelly-Massicotte}}, \bibinfo {author} {\bibfnamefont {A.}~\bibnamefont {Fedrizzi}},\ and\ \bibinfo {author} {\bibfnamefont {A.~M.}\ \bibnamefont {Brańczyk}},\ }\bibfield  {title} {\bibinfo {title} {Design considerations for high-purity heralded single-photon sources},\ }\href {https://doi.org/10.1103/PhysRevA.98.053811} {\bibfield  {journal} {\bibinfo  {journal} {Physical Review A}\ }\textbf {\bibinfo {volume} {98}},\ \bibinfo {pages} {053811} (\bibinfo {year} {2018})}\BibitemShut {NoStop}%
	\bibitem [{\citenamefont {Jabir}\ and\ \citenamefont {Samanta}(2017)}]{jabir_robust_2017}%
	\BibitemOpen
	\bibfield  {author} {\bibinfo {author} {\bibfnamefont {M.~V.}\ \bibnamefont {Jabir}}\ and\ \bibinfo {author} {\bibfnamefont {G.~K.}\ \bibnamefont {Samanta}},\ }\bibfield  {title} {\bibinfo {title} {Robust, high brightness, degenerate entangled photon source at room temperature},\ }\href {https://doi.org/10.1038/s41598-017-12709-5} {\bibfield  {journal} {\bibinfo  {journal} {Scientific Reports}\ }\textbf {\bibinfo {volume} {7}},\ \bibinfo {pages} {12613} (\bibinfo {year} {2017})}\BibitemShut {NoStop}%
	\bibitem [{\citenamefont {Meraner}\ \emph {et~al.}(2021)\citenamefont {Meraner}, \citenamefont {Chapman}, \citenamefont {Frick}, \citenamefont {Keil}, \citenamefont {Prilmüller},\ and\ \citenamefont {Weihs}}]{meraner_approaching_2021}%
	\BibitemOpen
	\bibfield  {author} {\bibinfo {author} {\bibfnamefont {S.}~\bibnamefont {Meraner}}, \bibinfo {author} {\bibfnamefont {R.}~\bibnamefont {Chapman}}, \bibinfo {author} {\bibfnamefont {S.}~\bibnamefont {Frick}}, \bibinfo {author} {\bibfnamefont {R.}~\bibnamefont {Keil}}, \bibinfo {author} {\bibfnamefont {M.}~\bibnamefont {Prilmüller}},\ and\ \bibinfo {author} {\bibfnamefont {G.}~\bibnamefont {Weihs}},\ }\bibfield  {title} {\bibinfo {title} {Approaching the {Tsirelson} bound with a {Sagnac} source of polarization-entangled photons},\ }\href {https://doi.org/10.21468/SciPostPhys.10.1.017} {\bibfield  {journal} {\bibinfo  {journal} {SciPost Physics}\ }\textbf {\bibinfo {volume} {10}},\ \bibinfo {pages} {017} (\bibinfo {year} {2021})}\BibitemShut {NoStop}%
	\bibitem [{\citenamefont {Xue}\ \emph {et~al.}(2021)\citenamefont {Xue}, \citenamefont {Niu}, \citenamefont {Liu}, \citenamefont {Duan}, \citenamefont {Chen}, \citenamefont {Pan}, \citenamefont {Jia}, \citenamefont {Wang}, \citenamefont {Liu}, \citenamefont {Zhang}, \citenamefont {Xu}, \citenamefont {Zhao}, \citenamefont {Cai}, \citenamefont {Gong}, \citenamefont {Hu}, \citenamefont {Xie},\ and\ \citenamefont {Zhu}}]{xue_ultrabright_2021}%
	\BibitemOpen
	\bibfield  {author} {\bibinfo {author} {\bibfnamefont {G.-T.}\ \bibnamefont {Xue}}, \bibinfo {author} {\bibfnamefont {Y.-F.}\ \bibnamefont {Niu}}, \bibinfo {author} {\bibfnamefont {X.}~\bibnamefont {Liu}}, \bibinfo {author} {\bibfnamefont {J.-C.}\ \bibnamefont {Duan}}, \bibinfo {author} {\bibfnamefont {W.}~\bibnamefont {Chen}}, \bibinfo {author} {\bibfnamefont {Y.}~\bibnamefont {Pan}}, \bibinfo {author} {\bibfnamefont {K.}~\bibnamefont {Jia}}, \bibinfo {author} {\bibfnamefont {X.}~\bibnamefont {Wang}}, \bibinfo {author} {\bibfnamefont {H.-Y.}\ \bibnamefont {Liu}}, \bibinfo {author} {\bibfnamefont {Y.}~\bibnamefont {Zhang}}, \bibinfo {author} {\bibfnamefont {P.}~\bibnamefont {Xu}}, \bibinfo {author} {\bibfnamefont {G.}~\bibnamefont {Zhao}}, \bibinfo {author} {\bibfnamefont {X.}~\bibnamefont {Cai}}, \bibinfo {author} {\bibfnamefont {Y.-X.}\ \bibnamefont {Gong}}, \bibinfo {author} {\bibfnamefont {X.}~\bibnamefont {Hu}}, \bibinfo {author} {\bibfnamefont {Z.}~\bibnamefont {Xie}},\ and\ \bibinfo {author}
		{\bibfnamefont {S.}~\bibnamefont {Zhu}},\ }\bibfield  {title} {\bibinfo {title} {Ultrabright {Multiplexed} {Energy}-{Time}-{Entangled} {Photon} {Generation} from {Lithium} {Niobate} on {Insulator} {Chip}},\ }\href {https://doi.org/10.1103/PhysRevApplied.15.064059} {\bibfield  {journal} {\bibinfo  {journal} {Physical Review Applied}\ }\textbf {\bibinfo {volume} {15}},\ \bibinfo {pages} {064059} (\bibinfo {year} {2021})}\BibitemShut {NoStop}%
	\bibitem [{\citenamefont {Krasnokutska}\ \emph {et~al.}(2019)\citenamefont {Krasnokutska}, \citenamefont {Chapman}, \citenamefont {Tambasco},\ and\ \citenamefont {Peruzzo}}]{krasnokutska_high_2019}%
	\BibitemOpen
	\bibfield  {author} {\bibinfo {author} {\bibfnamefont {I.}~\bibnamefont {Krasnokutska}}, \bibinfo {author} {\bibfnamefont {R.~J.}\ \bibnamefont {Chapman}}, \bibinfo {author} {\bibfnamefont {J.-L.~J.}\ \bibnamefont {Tambasco}},\ and\ \bibinfo {author} {\bibfnamefont {A.}~\bibnamefont {Peruzzo}},\ }\bibfield  {title} {\bibinfo {title} {High coupling efficiency grating couplers on lithium niobate on insulator},\ }\href {https://doi.org/10.1364/OE.27.017681} {\bibfield  {journal} {\bibinfo  {journal} {Optics Express}\ }\textbf {\bibinfo {volume} {27}},\ \bibinfo {pages} {17681} (\bibinfo {year} {2019})}\BibitemShut {NoStop}%
	\bibitem [{\citenamefont {Chen}\ \emph {et~al.}(2022)\citenamefont {Chen}, \citenamefont {Ruan}, \citenamefont {Fan}, \citenamefont {Wang}, \citenamefont {Liu}, \citenamefont {Li}, \citenamefont {Chen},\ and\ \citenamefont {Liu}}]{chen_low-loss_2022}%
	\BibitemOpen
	\bibfield  {author} {\bibinfo {author} {\bibfnamefont {B.}~\bibnamefont {Chen}}, \bibinfo {author} {\bibfnamefont {Z.}~\bibnamefont {Ruan}}, \bibinfo {author} {\bibfnamefont {X.}~\bibnamefont {Fan}}, \bibinfo {author} {\bibfnamefont {Z.}~\bibnamefont {Wang}}, \bibinfo {author} {\bibfnamefont {J.}~\bibnamefont {Liu}}, \bibinfo {author} {\bibfnamefont {C.}~\bibnamefont {Li}}, \bibinfo {author} {\bibfnamefont {K.}~\bibnamefont {Chen}},\ and\ \bibinfo {author} {\bibfnamefont {L.}~\bibnamefont {Liu}},\ }\bibfield  {title} {\bibinfo {title} {Low-loss fiber grating coupler on thin film lithium niobate platform},\ }\href {https://doi.org/10.1063/5.0093033} {\bibfield  {journal} {\bibinfo  {journal} {APL Photonics}\ }\textbf {\bibinfo {volume} {7}},\ \bibinfo {pages} {076103} (\bibinfo {year} {2022})}\BibitemShut {NoStop}%
	\bibitem [{\citenamefont {Yu}\ \emph {et~al.}(2023)\citenamefont {Yu}, \citenamefont {Zhong}, \citenamefont {Fang}, \citenamefont {Patel}, \citenamefont {Li}, \citenamefont {Liu}, \citenamefont {Li}, \citenamefont {Xu}, \citenamefont {Sagona-Stophel}, \citenamefont {Mer}, \citenamefont {Thomas}, \citenamefont {Meng}, \citenamefont {Li}, \citenamefont {Yang}, \citenamefont {Wang}, \citenamefont {Guo}, \citenamefont {Zhang}, \citenamefont {Tranmer}, \citenamefont {Dong}, \citenamefont {Wang}, \citenamefont {Tang}, \citenamefont {Li}, \citenamefont {Walmsley},\ and\ \citenamefont {Guo}}]{yu_universal_2023}%
	\BibitemOpen
	\bibfield  {author} {\bibinfo {author} {\bibfnamefont {S.}~\bibnamefont {Yu}}, \bibinfo {author} {\bibfnamefont {Z.-P.}\ \bibnamefont {Zhong}}, \bibinfo {author} {\bibfnamefont {Y.}~\bibnamefont {Fang}}, \bibinfo {author} {\bibfnamefont {R.~B.}\ \bibnamefont {Patel}}, \bibinfo {author} {\bibfnamefont {Q.-P.}\ \bibnamefont {Li}}, \bibinfo {author} {\bibfnamefont {W.}~\bibnamefont {Liu}}, \bibinfo {author} {\bibfnamefont {Z.}~\bibnamefont {Li}}, \bibinfo {author} {\bibfnamefont {L.}~\bibnamefont {Xu}}, \bibinfo {author} {\bibfnamefont {S.}~\bibnamefont {Sagona-Stophel}}, \bibinfo {author} {\bibfnamefont {E.}~\bibnamefont {Mer}}, \bibinfo {author} {\bibfnamefont {S.~E.}\ \bibnamefont {Thomas}}, \bibinfo {author} {\bibfnamefont {Y.}~\bibnamefont {Meng}}, \bibinfo {author} {\bibfnamefont {Z.-P.}\ \bibnamefont {Li}}, \bibinfo {author} {\bibfnamefont {Y.-Z.}\ \bibnamefont {Yang}}, \bibinfo {author} {\bibfnamefont {Z.-A.}\ \bibnamefont {Wang}}, \bibinfo {author} {\bibfnamefont {N.-J.}\ \bibnamefont {Guo}}, \bibinfo
		{author} {\bibfnamefont {W.-H.}\ \bibnamefont {Zhang}}, \bibinfo {author} {\bibfnamefont {G.~K.}\ \bibnamefont {Tranmer}}, \bibinfo {author} {\bibfnamefont {Y.}~\bibnamefont {Dong}}, \bibinfo {author} {\bibfnamefont {Y.-T.}\ \bibnamefont {Wang}}, \bibinfo {author} {\bibfnamefont {J.-S.}\ \bibnamefont {Tang}}, \bibinfo {author} {\bibfnamefont {C.-F.}\ \bibnamefont {Li}}, \bibinfo {author} {\bibfnamefont {I.~A.}\ \bibnamefont {Walmsley}},\ and\ \bibinfo {author} {\bibfnamefont {G.-C.}\ \bibnamefont {Guo}},\ }\bibfield  {title} {\bibinfo {title} {A universal programmable {Gaussian} boson sampler for drug discovery},\ }\href {https://doi.org/10.1038/s43588-023-00526-y} {\bibfield  {journal} {\bibinfo  {journal} {Nature Computational Science}\ }\textbf {\bibinfo {volume} {3}},\ \bibinfo {pages} {839} (\bibinfo {year} {2023})}\BibitemShut {NoStop}%
	\bibitem [{\citenamefont {Huh}\ \emph {et~al.}(2015)\citenamefont {Huh}, \citenamefont {Guerreschi}, \citenamefont {Peropadre}, \citenamefont {McClean},\ and\ \citenamefont {Aspuru-Guzik}}]{huh_boson_2015}%
	\BibitemOpen
	\bibfield  {author} {\bibinfo {author} {\bibfnamefont {J.}~\bibnamefont {Huh}}, \bibinfo {author} {\bibfnamefont {G.~G.}\ \bibnamefont {Guerreschi}}, \bibinfo {author} {\bibfnamefont {B.}~\bibnamefont {Peropadre}}, \bibinfo {author} {\bibfnamefont {J.~R.}\ \bibnamefont {McClean}},\ and\ \bibinfo {author} {\bibfnamefont {A.}~\bibnamefont {Aspuru-Guzik}},\ }\bibfield  {title} {\bibinfo {title} {Boson sampling for molecular vibronic spectra},\ }\href {https://doi.org/10.1038/nphoton.2015.153} {\bibfield  {journal} {\bibinfo  {journal} {Nature Photonics}\ }\textbf {\bibinfo {volume} {9}},\ \bibinfo {pages} {615} (\bibinfo {year} {2015})}\BibitemShut {NoStop}%
	\bibitem [{\citenamefont {Sempere-Llagostera}\ \emph {et~al.}(2022)\citenamefont {Sempere-Llagostera}, \citenamefont {Patel}, \citenamefont {Walmsley},\ and\ \citenamefont {Kolthammer}}]{sempere-llagostera_experimentally_2022}%
	\BibitemOpen
	\bibfield  {author} {\bibinfo {author} {\bibfnamefont {S.}~\bibnamefont {Sempere-Llagostera}}, \bibinfo {author} {\bibfnamefont {R.}~\bibnamefont {Patel}}, \bibinfo {author} {\bibfnamefont {I.}~\bibnamefont {Walmsley}},\ and\ \bibinfo {author} {\bibfnamefont {W.}~\bibnamefont {Kolthammer}},\ }\bibfield  {title} {\bibinfo {title} {Experimentally {Finding} {Dense} {Subgraphs} {Using} a {Time}-{Bin} {Encoded} {Gaussian} {Boson} {Sampling} {Device}},\ }\href {https://doi.org/10.1103/PhysRevX.12.031045} {\bibfield  {journal} {\bibinfo  {journal} {Physical Review X}\ }\textbf {\bibinfo {volume} {12}},\ \bibinfo {pages} {031045} (\bibinfo {year} {2022})}\BibitemShut {NoStop}%
	\bibitem [{\citenamefont {Silverstone}\ \emph {et~al.}(2016)\citenamefont {Silverstone}, \citenamefont {Bonneau}, \citenamefont {O’Brien},\ and\ \citenamefont {Thompson}}]{silverstone_silicon_2016}%
	\BibitemOpen
	\bibfield  {author} {\bibinfo {author} {\bibfnamefont {J.~W.}\ \bibnamefont {Silverstone}}, \bibinfo {author} {\bibfnamefont {D.}~\bibnamefont {Bonneau}}, \bibinfo {author} {\bibfnamefont {J.~L.}\ \bibnamefont {O’Brien}},\ and\ \bibinfo {author} {\bibfnamefont {M.~G.}\ \bibnamefont {Thompson}},\ }\bibfield  {title} {\bibinfo {title} {Silicon {Quantum} {Photonics}},\ }\href {https://doi.org/10.1109/JSTQE.2016.2573218} {\bibfield  {journal} {\bibinfo  {journal} {IEEE Journal of Selected Topics in Quantum Electronics}\ }\textbf {\bibinfo {volume} {22}},\ \bibinfo {pages} {390} (\bibinfo {year} {2016})}\BibitemShut {NoStop}%
	\bibitem [{\citenamefont {Saravi}\ \emph {et~al.}(2021)\citenamefont {Saravi}, \citenamefont {Pertsch},\ and\ \citenamefont {Setzpfandt}}]{saravi_lithium_2021}%
	\BibitemOpen
	\bibfield  {author} {\bibinfo {author} {\bibfnamefont {S.}~\bibnamefont {Saravi}}, \bibinfo {author} {\bibfnamefont {T.}~\bibnamefont {Pertsch}},\ and\ \bibinfo {author} {\bibfnamefont {F.}~\bibnamefont {Setzpfandt}},\ }\bibfield  {title} {\bibinfo {title} {Lithium {Niobate} on {Insulator}: {An} {Emerging} {Platform} for {Integrated} {Quantum} {Photonics}},\ }\href {https://doi.org/10.1002/adom.202100789} {\bibfield  {journal} {\bibinfo  {journal} {Advanced Optical Materials}\ }\textbf {\bibinfo {volume} {9}},\ \bibinfo {pages} {2100789} (\bibinfo {year} {2021})}\BibitemShut {NoStop}%
\end{thebibliography}
%

\clearpage
\section*{Supplementary Material}
\renewcommand\thefigure{S\arabic{figure}}    
\setcounter{figure}{0}

\section{Effects impacting fidelity}

We explore several factors that can impact the fidelity of the generated $N00N$ state, and the visibility of the measured quantum interference.

\subsection{Input state balance}
 
\begin{figure}
    \includegraphics[width=1\linewidth]{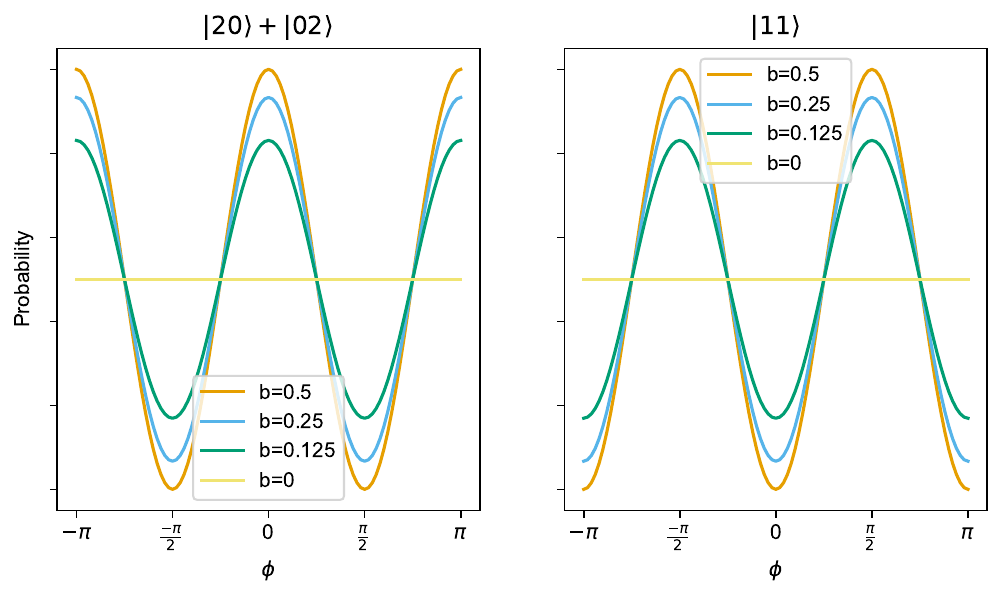}
    \caption{Impact of the $N00N$ state balance on the probability of photon bunching or anti-bunching.}
    \label{fig:sup_balance}
\end{figure}

We simulate the impact of an imbalanced $N00N$ state, given as
\begin{equation}
    \ket{\psi} = \sqrt{b}\ket{2_a0_b} + e^{2i\phi}\sqrt{1-b} \ket{0_a2_b},
\end{equation}
where the parameter $b$ controls the balance between the two sources.
$b=0$ is equivalent to pumping a single SPDC source and $b=0.5$ is when a balance $N00N$ state is generated.
Figure S1 shows the output probability of measuring two photons in a single output port $|\braket{2_a0_b|\psi}|^2 + |\braket{0_a2_b|\psi}|^2$ and the output probability for measuring $|\braket{1_a1_b|\psi}|^2$, as a function of the phase $\phi$.

\subsection{Quantum state purity}

We next simulate the impact of indistinguishability between the two sources as a reduction in the purity of the quantum state. 
We model this as decoherence of the two photon $N00N$ state with the density matrix
\begin{equation}
    \rho = \begin{pmatrix}
    0.5 & 0 & 0 & 0.5p \\
    0 & 0 & 0 & 0 \\
    0 & 0 & 0 & 0 \\
    0.5p & 0 & 0 & 0.5
\end{pmatrix},
\end{equation}
where $p$ is the purity of the state, which here is equivalent to the spectral overlap of the two sources.
We assume here that the two sources are balanced in generation efficiency.
Figure S2 shows the output probability of the two photons exiting the same port and of exiting in different ports as a function of the phase $\phi$.
As the purity decreases, the interference visibility also degrades.

\begin{figure}
    \centering
    \includegraphics[width=1\linewidth]{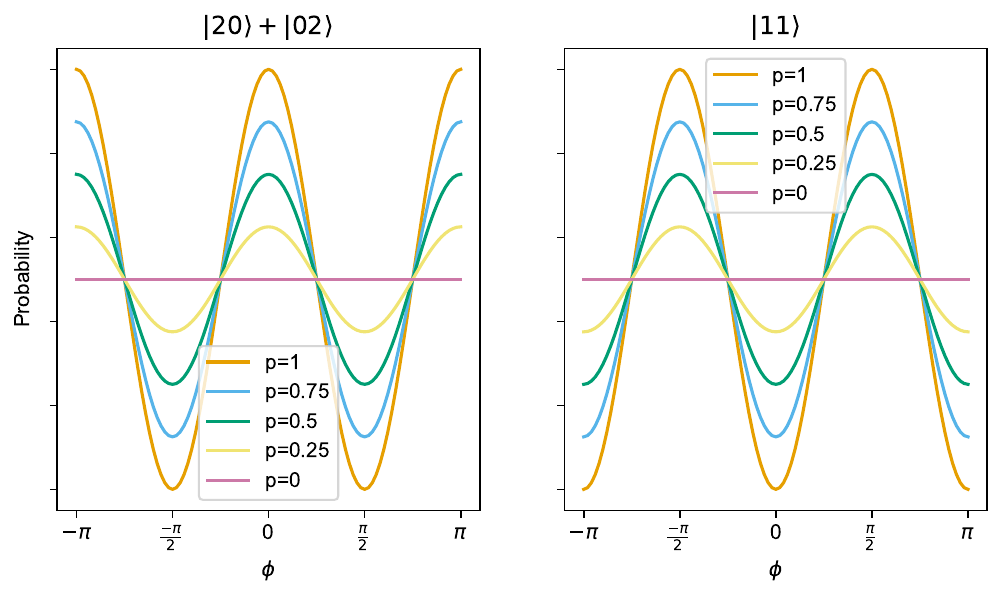}
    \caption{Impact of the purity of the $N00N$ state on the probability of photon bunching or anti-bunching.}
    \label{fig:sup_purity}
\end{figure}

\section{Periodically poling}

We first deposit comb-shaped chrome electrodes on the surface of the lithium niobate-on-insulator wafers. 
The poling electrodes have a period of \SI{2.478}{\um}, a duty cycle of \SI{25}{\percent}, a length of \SI{2}{mm} and a gap of \SI{30}{\um}.
We next apply a series of thirty \SI{670}{V} \SI{1}{ms}-long electrical pulses to invert the crystal domain between the electrodes.
We confirm the successful poling using a two-photon microscope, as shown in Figure S3.

\begin{figure}[h]
    \centering
    \includegraphics[width=0.75\linewidth]{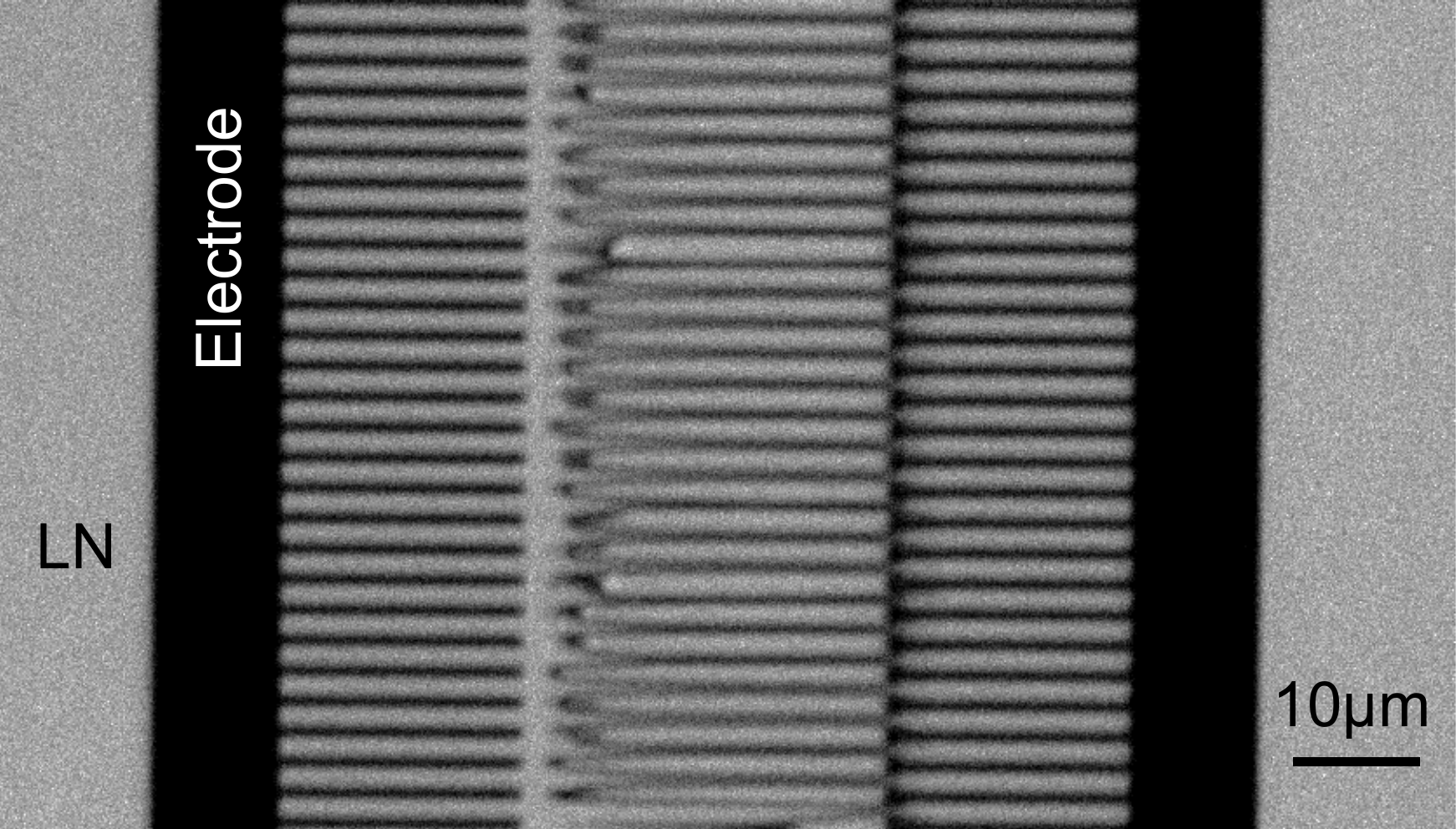}
    \caption{Two photon microscope image of a periodically poled region of LNOI including the electrodes. The period is \SI{2.478}{\um} and gap is \SI{30}{\um}}
    \label{fig:sup_poling}
\end{figure}

This gap is sufficient to fit multiple waveguides for several SPDC sources on a single device. 
Increasing the separation to \SI{75}{\um}, as was shown in \cite{wang_ultrahigh-efficiency_2018}, would enable 10 waveguides with \SI{5}{\um} gaps, which is sufficient to have negligible evanescent coupling.
Employing adaptive poling, where the periodic is varied according to the lithium niobate film thickness could be used to have multiple poling regions with identical phase matching \cite{chen_adapted_2024}.

\end{document}